\documentclass[12pt]{article}

\usepackage[colorlinks,linkcolor=blue,citecolor=blue,urlcolor=blue,bookmarks,bookmarksnumbered]{hyperref}
\usepackage{amsmath,amssymb,amsfonts}
\usepackage[paper=letterpaper,margin=1.0in]{geometry}
\usepackage{graphicx,xcolor}
\usepackage{braket}
\usepackage{caption}
\usepackage{subcaption}
\usepackage{cite}
\usepackage{mathrsfs}

\newcommand{\be}{\begin{equation}}
\newcommand{\ee}{\end{equation}}
\newcommand{\bea}{\begin{eqnarray}}
\newcommand{\eea}{\end{eqnarray}}
\newcommand{\ba}{\begin{array}}
\newcommand{\ea}{\end{array}}
\newcommand{\ben}{\begin{enumerate}}
\newcommand{\een}{\end{enumerate}}
\newcommand{\bi}{\begin{itemize}}
\newcommand{\ei}{\end{itemize}}
\newcommand{\bc}{\begin{center}}
\newcommand{\ec}{\end{center}}
\newcommand{\bfig}{\begin{figure}}
\newcommand{\efig}{\end{figure}}
\newcommand{\bq}{\begin{quotation}}
\newcommand{\eq}{\end{quotation}}
\newcommand{\bt}{\begin{table}}
\newcommand{\et}{\end{table}}
\newcommand{\btab}{\begin{tabular}}
\newcommand{\etab}{\end{tabular}}
\newcommand{\bs}{\begin{slide}}
\newcommand{\es}{\end{slide}}

\newcommand{\pa}{\partial}

\newcommand{\IR}{\mathbb{R}}

\newcommand{\X}{\mathbb{X}}

\def\pa{\partial}

\def\S{\mathbb{S}}

\def\S{\mathbb{S}}

\let\ba=\overline

\let\d=\delta

\def\rd{{\rm d}}

\let\g=\gamma

\let\j=\psi

\let\w=\omega

\def\IR{\relax\leavevmode{\rm I\kern-.18em R}}
\def\ZZ{\relax\leavevmode
       \ifmmode\mathchoice
       {\hbox{\sf Z\kern-.4em Z}}
       {\hbox{\sf Z\kern-.4em Z}}
       {\lower.9pt\hbox{\scriptsize\sf Z\kern-.36em Z}}
       {\lower1.2pt\hbox{\tiny\sf Z\kern-.36em Z}}
       \else{\sf Z\kern-.4em Z}\fi}
\def\RR{\relax\leavevmode
       \ifmmode\mathchoice
       {\hbox{\sf R\kern-.4em R}}
       {\hbox{\sf R\kern-.4em R}}
       {\lower.9pt\hbox{\scriptsize\sf R\kern-.36em R}}
       {\lower1.2pt\hbox{\tiny\sf R\kern-.36em R}}
       \else{\sf R\kern-.4em R}\fi}

\def\resetby#1#2{\@addtoreset{#2}{#1}}
\def\seceq{\@addtoreset{equation}{section}
              \def\theequation{\thesection.\arabic{equation}}}

\def\Label#1{\label{#1}%
                \smash{\hbox to0pt{\raise1ex\hbox{\tiny[#1]}\hss}}}
\def\noLabels{\let\Label=\label}

\makeatletter
\DeclareRobustCommand\widecheck[1]{{\mathpalette\@widecheck{#1}}}
\def\@widecheck#1#2{%
    \setbox\z@\hbox{\m@th$#1#2$}%
    \setbox\tw@\hbox{\m@th$#1%
       \widehat{%
          \vrule\@width\z@\@height\ht\z@
          \vrule\@height\z@\@width\wd\z@}$}%
    \dp\tw@-\ht\z@
    \@tempdima\ht\z@ \advance\@tempdima2\ht\tw@ \divide\@tempdima\thr@@
    \setbox\tw@\hbox{%
       \raise\@tempdima\hbox{\scalebox{1}[-1]{\lower\@tempdima\box\tw@}}}%
    {\ooalign{\box\tw@ \cr \box\z@}}}
\makeatother

 \allowdisplaybreaks

\begin{document}

{\footnotesize
${}$
}

\bc

\vskip 1.0cm

{\Large \bf Triple Interference, Non-linear Talbot Effect}\\
\vskip 0.1cm
{\Large \bf and Gravitization of the Quantum}\\

\vskip 1.0cm

\renewcommand{\thefootnote}{\fnsymbol{footnote}}

\bf Per Berglund${}^{1}$\footnote{Per.Berglund@unh.edu},  
\bf Andrew Geraci${}^{2}$\footnote{andrew.geraci@northwestern.edu}, 
\bf Tristan H{\"u}bsch${}^{3}$\footnote{thubsch@howard.edu}
{\bf  David Mattingly${}^{1}$\footnote{Corresponding author: david.mattingly@unh.edu} and} 
{ \bf Djordje Minic${}^{4}$\footnote{dminic@vt.edu} } \\

\vskip 0.5cm

{\it
${}^1$Department of Physics and Astronomy, University of New Hampshire, Durham, NH 03824, U.S.A. \\
${}^2$Department of Physics and Astronomy, Northwestern University, Evanston, IL 60208, U.S.A. \\
${}^3$Department of Physics and Astronomy, Howard University, Washington, DC 20059, U.S.A. \\
${}^4$Department  of Physics, Virginia Tech, Blacksburg, VA 24061, U.S.A. \\
}

\ec

\vskip 1.0cm

\begin{abstract}
Recently we have discussed a new approach to the problem of quantum gravity in which
the quantum mechanical structures that are traditionally fixed, such as the Fubini-Study metric in the Hilbert space of states, become dynamical and so implement the idea of { gravitizing the quantum}.  
In this paper we elaborate on a specific test of this new approach to quantum gravity using triple interference in a varying gravitational field.
Our discussion is driven by a profound analogy with recent triple-path interference experiments performed in the context of non-linear optics.
We emphasize that the triple interference experiment in a varying gravitational field would deeply influence the present understanding of
the kinematics of quantum gravity and quantum gravity phenomenology. We also discuss the non-linear Talbot effect as another striking phenomenological probe of  gravitization of the geometry of quantum theory.

\end{abstract}

\vspace{0.5cm}

\begin{center}

\end{center}

\vspace{0.5cm}

\renewcommand{\thefootnote}{\arabic{footnote}}

\newpage

\section{Introduction: What is quantum gravity?}
\label{s:Intro}

In\cite{Berglund:2022skk}, we have recently discussed a new approach to 
the problem of quantum gravity, which is still an outstanding fundamental question in physics.  In standard approaches to quantum gravity, one applies wholesale the structure of quantum mechanics to the gravitational field, modifying the geometrical and topological structures of general relativity as necessary. Whether the modifications are introducing extended objects and varying dimensions, as in string theory, or a different geometrical description of the gauge invariant reduced phase space as in loop quantum gravity, or different symmetries as in supergravity or Horava-Lifshitz gravity (among other approaches), quantum mechanics remains unchanged while gravity, space, and time get modified. Since quantization of the other three fundamental forces is well understood using this prescription, and since quantum mechanics is a fairly rigid structure, this approach has dominated the landscape of quantum gravity.
 
Fundamentally, however, general relativity contains as deep conceptual elements as quantum mechanics. Hence there is also the possibility to operate in reverse; rather than ``quantizing gravity'' the correct theory may instead ``gravitize the quantum'' and profoundly change the structure of quantum mechanics. 
In ~\cite{Berglund:2022skk, Berglund:2022qcc} we proposed one such possibility by
extending the dynamical aspects of general covariance, that is, the statement that all quantities in our physical theories must be dynamical, to usually fixed quantum mechanical structures. This approach was inspired by our previous work on {\em\/metastring\/} theory~\cite{Freidel:2015uug, Freidel:2016pls, Freidel:2013zga, Freidel:2014qna, Freidel:2015pka, Freidel:2017xsi,  Freidel:2018apz,  Freidel:2019jor, Minic:2020oho,  Freidel:2017wst, Freidel:2017nhg} 
and its various ramifications
\cite{Freidel:2021wpl, Gunaydin:2013nqa, Berglund:2020qcu} as well as work on the geometric aspects of quantum foundations\footnote{This proposal is similar to ideas that have been discussed by Penrose as a way to address the quantum measurement problem~\cite{penrose} 
and to attempts to understand deeper axiomatic foundations of quantum theory
and quantum gravity~\cite{Hardy:2001jk}.}~\cite{Minic:2002pd, Minic:2003en, Minic:2004rj,  Jejjala:2007rn, Minic:2020zjb}. 

General covariance can be split into two distinct ideas~\cite{Norton}: 1) coordinate invariance, which
can be applied to almost any physical theory, and 2) dynamicism, which is much more profound as it postulates that all quantities in our physical theories must be dynamical. It is dynamicism that takes us from the fixed stage of Newtonian gravity and special relativity into the full framework of general relativity.
In \cite{Berglund:2022skk}  we proposed that quantum geometry, that is, the actual geometry typically imposed on Hilbert spaces by quantum mechanical axioms and not a quantized spacetime geometry, becomes dynamical in quantum gravity in a way that is consistent with the principles of unitarity and causality. This proposal  sheds new light on the new field of quantum gravity phenomenology~\cite{Addazi:2021xuf}, which at present covers everything from searches for spacetime symmetry violation in high energy astrophysics to proposals that employ quantum information techniques and nanomechanical oscillators to show empirically that gravity is quantized\cite{Bose:2017nin, Marletto:2017kzi, Belenchia:2018szb,Carney:2022dku}.
In addition to these experimental searches we have proposed to test this new approach to quantum gravity by utilizing multi-path interference and optical lattice atomic clocks \cite{Berglund:2022skk}.

In this paper we elaborate on a specific test of this new approach to quantum gravity using triple interference in a varying gravitational field.  The key point is that if the geometric structures of quantum mechanics are gravitized and dynamical, then the metric on the Hilbert space, usually defined by the Born rule where the probability $P$ is given by $P=\delta_{ab} \psi_a \psi_b$, where $a,b$ are state space indices, must be generalized to be a function of the state space variables, or quantum state.  In other words $P=g_{ab}(\psi_{a,b,c,...}) \psi_a \psi_b$. Unavoidably, making the Hilbert space metric dynamical introduces non-linearities in the Born rule.  Without a specified ``gravitized quantum dynamics'', which is far beyond the scope of this paper, the phenomenology is dictated by a parameterization approach, where the possible non-linearities are parameterized and tested experimentally.  This is the approach we take in this paper, concentrating on the first possible non-linearity, that of triple-path interference.

Our presentation is driven by a profound analogy with recent triple-path interference experiments performed in the context of non-linear optics.
We emphasize that the triple interference experiment in a varying gravitational field would deeply influence the present understanding of
the kinematics of quantum gravity and quantum gravity phenomenology. In particular, the single Hilbert space picture with a rigid maximally symmetric geometry of the space of states, that is tied to the robustness of the Born rule, would have to be generalized to a higher categorical structure, with observer dependent Hilbert spaces. As a natural generalization of triple interference, we also discuss the non-linear Talbot effect as another remarkable 
phenomenological probe of
gravitization of the geometry of quantum theory.

The paper is organized as follows: In section~\ref{s:3+ference} we review the concept of triple, and higher order, interference in general.
Then in section~\ref{s:NLO} we discuss triple interference in the context of non-linear optics as a well-known analogy.
In section~\ref{s:3nterference} we propose a concrete model of triple interference in the context of a gravitized quantum theory and in section~\ref{s:dragons} we partially answer some general criticisms relating to possible modifications of quantum theory.
In section~\ref{s:GQ=QG}, we theoretically tie this proposal to both a theoretical proposal for quantum gravity as well as a geometric view of quantum theory.  The latter sections focus on experimental and phenomenological concerns.
In section~\ref{s:scale} we discuss the scale of triple interference involving quantum gravity.
In section~\ref{s:toy} we discuss a toy model for triple interference in quantum gravity and the modification of
the Born rule. Section~\ref{s:Talbot} discusses the non-linear Talbot effect with quantum matter waves in the context of quantum gravity, viewed as gravitization of quantum theory.
Finally, in section~\ref{s:Exp} we make some comments regarding the experimental set-up and collect some concluding thoughts in section~\ref{s:Coda}.

\section{Triple, and higher order, interference}
\label{s:3+ference}

As pointed out in our recent work  \cite{Berglund:2022skk} 
a very interesting conceptual  and experimental consequence of the gravitization of quantum theory would be
an intrinsic triple-path, and higher order, interference originally discussed 
in a different context by Sorkin~\cite{Sorkin:1994dt}.
Let us denote by
$
P_{n}(A,B,C,\cdots)
$
the probability of a system to go from an initial state $\ket{\alpha}$ to a final state $\ket{\beta}$
when $n$ paths $A,B,C,\dots$ connecting the two are available, following the presentation in~\cite{Huber:2021xpx}.
Classically, we have
\be
P_{n}(A,B,C,\cdots) \,=\, P_{1}(A) + P_{1}(B) + P_{1}(C) + \cdots\;,
\ee
for any number of paths.
Quantum mechanically, we have for two paths
$
P_{2}(A,B) = |\psi_A + \psi_B|^2 \vphantom{\Big|}$
or more explicitly
\be
{|\psi_A|^2} + 
{|\psi_B|^2} + 
{(\psi_A^*\psi_B^{\phantom{*}} + \psi_B^*\psi_A^{\phantom{*}})}
\equiv P_{1}(A) + P_{1}(B) + I_{2}(A,B)
\ee
where the last term
\be
I_{2}(A,B) = P_{2}(A,B)-P_{1}(A)-P_{1}(B) 
\ee
is the ``interference'' of the two paths $A$ and $B$.
The non-vanishing of this double-path interference, 
$I_{2}(A,B)\neq 0$, distinguishes quantum theory from the classical one.
The Born rule dictates that all the superimposed paths only interfere with each other 
in a pairwise manner.
For instance, for three paths we have
\be
P_{3}(A,B,C) = |\psi_A {+} \psi_B {+} \psi_C|^2 \equiv  P_{2}(A,B) {+} P_{2}(B,C) {+} P_{2}(C,A) 
{-} P_{1}(A) {-} P_{1}(B) {-} P_{1}(C),
\label{eq:three-slit}
\ee
where only pairwise interferences between the pairs $(A,B)$, $(B,C)$, and $(C,A)$ appear.

It is clear from the above that in order for there to be a non-linear correction in an interference pattern the Born rule must be relaxed.  
In particular, the Born rule itself should be modified by additional non-linear terms when calculating probabilities (cf.~\cite{Helou and Chen}).  The modified Born rule would then generate deviations from the expected interference patterns in quantum mechanics.

Sorkin has discussed intrinsic higher order interferences in~\cite{Sorkin:1994dt}.  Consider a triple slit experiment
as in equation~\eqref{eq:three-slit}.
Since only pairwise interferences between the pairs $(A,B)$, $(B,C)$, and $(C,A)$ appear, it makes sense to define any deviation from this relation as the intrinsic
triple-path interference~\cite{Sorkin:1994dt} 
\be
I_{3}(A,B,C) \overset{\scriptscriptstyle\text{def}}=
P_{3}(A,B,C)
-P_{2}(A,B)
-P_{2}(B,C)
-P_{2}(C,A)
+P_{1}(A)
+P_{1}(B)
+P_{1}(C).
\label{e:I3}
\ee
(This can be easily generalized for the case of $n$-paths.)
For both classical and quantum theory, this intrinsic triple-path interference is zero for any triplet of
paths. Experimental confirmation of $I_3=0$ would be a confirmation of the Born rule.  As discussed in~\cite{Huber:2021xpx} in Refs.~\cite{Sinha,Park}, bounds were placed on the parameter
\be
\kappa = \dfrac{\varepsilon}{\delta}, \quad \varepsilon =  I_3(A,B,C), \quad \delta  =  |I_2(A,B)| + |I_2(B,C)| + |I_2(C,A)|.
\ee
Ref.~\cite{Sinha} reports $\kappa=0.0064\pm 0.0119$ for a multi-slit experiment
with a single photon source, while Ref.~\cite{Park} reports $\kappa=0.007\pm 0.003$
based on a liquid state NMR experiment.
Thus, the 1$\sigma$ deviation of $\kappa$ from zero allowed by these experiments
is $|\kappa| < 0.01\sim 0.02$. (See also~\cite{Lutz1, Lutz2}.) Quite surprisingly~\cite{Huber:2021xpx}, the neutrino test of triple-interference based on the
forthcoming JUNO experiment gives the same order of precision as the electromagnetic tests.

In \cite{Berglund:2022skk} we have suggested to generalize Sorkin's approach to the case of gravity.  In particular, one can consider triple-slit interference experiments (either involving photons in a three-slit experiment or 
oscillations of the three neutrino flavors) in a gravitational field. These experiments might also call for the use of gravitational or atomic interferometry as well as optical lattice atomic clocks.
The intrinsic triple correlator not present in the canonical quantum theory should be present to some degree given the above general argument presented in \cite{Berglund:2022skk}. In fact, the natural conjecture here is that due to the intrinsic non-linearity of
gravity one should, in principle, expect interferences of all orders in the tower of Sorkin like interferences, and also a non-linear
summation of these. 

Before we proceed with the details of our proposal, we must note that seeming violations of the Born rule have occurred in other contexts.  A well-known example is the apparent deviation from the Born rule that occurs in a triple slit experiment when one considers non-classical paths~\cite{nonclassical}. In this scenario the experimentalist runs multiple experiments. First they run a three slit experiment, then the various two slit and one slit variations.  If the Born rule holds \textit{and} one assumes that the total probabilities for two/three slit propagation are given simply by the sum of the single slit amplitudes, i.e. $\psi_{ABC}=\psi_A + \psi_B + \psi_C$ then Eq.~(\ref{e:I3}) holds.  The apparent violation of this rule comes when one considers non-classical paths, i.e. paths that loop through multiple slits.  Due to the existence of these paths, which are only possible when there are multiple slits, the second assumption above, that one can just add wavefunctions to get the three slit amplitude, fails.  Hence there will be an apparent violation of the Born rule.

This phenomenon, while couched in the language of quantum foundations, is actually familiar in many contexts.  Consider for example an experiment with an emitter and a charge detector that detects whether a charged particle has passed by.  We can set up two initial options, just as in a double slit experiment: we can launch a single proton or electron from an emitter and the detector will register each outcome as a click.  Now we could ask what happens if we allow for both options, where we emit both an electron and a proton.  Is the total wavefunction and measurement output then simply the sum of the two?  In the non-interacting case the answer is of course yes, the charge detector will click twice.  In the interacting case the answer is clearly no, there is also the possibility that they will form a neutral bound state and the detector will not register the emitted particles.  This can be considered as an extreme violation of the Born rule - the individual free particle scenarios do not just add to make the joint scenario.  We of course don't interpret it as a fundamental violation of the Born rule, we understand the existence of the bound state needs to be added into our path integral.  However, the effective violation of Born approach is precisely used in the form of ``missing matter'' or energy experiments at particle accelerators --- the loss of energy or particle number during a scattering experiment, which is an apparent violation of Born, is evidence that there are new bound or particle states present that need to be added into the path integral.

In both the non-classical paths and the scattering experiments, the apparent failure of Born is due to new paths and states that do not appear when considering only parts of the Hilbert space independently.  In the non-classical paths scenario, these are the paths that loop through multiple slits.  In the scattering scenario it is bound states that aren't representable as a sum of states in the individual free particle Hilbert spaces.  In neither case is the Born rule actually modified and the apparent failure comes simply from the fact that the full Hilbert space isn't factorizable into the sub-Hilbert spaces that represent the outcomes of the simpler experiments.  Born rule modifications can \textit{only} come from modifying the measure of the path integral in quantum mechanics.  Neglecting these extra states in both cases can be re-formulated as a modification of the measure --- one is arbitrarily setting some allowed paths to not contribute, i.e. have measure zero in the path integral.  This gives an effective modification and Born violation, effectively due to ignorance of part of the allowed phase space.
In contrast, our proposal intrinsically modifies the Born rule for \textit{any} experiment that we conceive of that occurs in classical spacetime.

\section{Triple interference and non-linear optics}
\label{s:NLO}
Intrinsic triple interference has been recently discussed in the context of non-linear optics.
In this section we review the recent discussion  \cite{Namdar:2021czo} of this topic.
We emphasize that in this particular discussion one does not have a fundamental generalization of Born's rule.
Non-linear optics is an effective classical description of phenomena that are essentially quantum, and described
by quantum optics and quantum electrodynamcs, both of which obey the fundamental postulates of quantum theory, including the Born rule. However, the analogy is very instructive and will pave the way for the discussion of triple interference in the context of quantum gravity in section~\ref{s:3nterference}.

Following the exposition in Thorne and Blandford \cite{mcp}, the necessary equations for non-linear optics are as follows.  The polarization $\mathscr{P}_i$ ($i=1,2,3$) is a general non-linear function of the electric field $E_i$
\be
\mathscr{P}_i = a_{ij} E_j + b_{ijk} E_j E_k+...
\ee
(The ``cubic'' coefficients $b_{ijk}$ are described in \cite{mcp}.)
The energy associated with this expression 
(the  measurable quantity in non-linear optics that translates into generalized probability in the quantum gravity context)
is essentially given 
as the scalar product of the polarization and electric field vectors
\be
U =  a_{ij} E_i E_j + b_{ijk} E_i E_j E_k +...
\label{e:U}
\ee
where $i,j,k$ are spatial indices $i, j, k = 1,2,3$.
Next, one represents the electric field $E_i$ in terms of a complex quantity that involves the amplitude $A_i$ and the phase determined by the
frequency $\omega$ and the wave vector $k_i$.
One uses the Maxwell equations in a medium determined by the above non-linear relation between $\mathscr{P}_i$ and $E_i$,
and to the leading order in the gradient expansion (the specific assumptions are spelled out in chapter 10 of \cite{mcp})
and the dependence of amplitude of one direction $z$,
one obtains the following {\it schematic} fundamental formula
governing the non-linear evolution (given the conservation of energy/frequency, $\omega_1 = \omega_2 + \omega_3$,
and momenta/wave vectors, $k_{1i} = k_{2i} + k_{3i}$)
\be
\frac{\rd A_i}{\rd z} = b_{ijk} A_i A_j.
\label{e:dA=AA}
\ee
(We will be more specific about this equation in what follows because the complex conjugation is involved in general.)
This equation describes the triple-frequency mixing, which leads to triple interference and therefore a non-zero $I_3$ from
the previous section.

Let us be a bit more specific \cite{mcp}:
Suppose the two waves $n=1$ and $n=2$ are coupled in an 
anisotropic crystal.
Also, let the corresponding electric field vectors be given by the real part of \cite{mcp}
\be
E_j^{(n)} = A_j^{(n)} \exp[i({\vec{k}}_n \cdot \vec{x} - \omega_n t )].
\ee
These waves couple through the second-order nonlinear (``cubic'') susceptibility $b_{ijk}$
and give rise to the non-linear polarization vector \cite{mcp}
\be
\mathscr{P}_i^{(3)} = b_{ijk} E_{j}^{(1)} E_{k}^{(2)}.
\ee
This in turn generates the 3-rd component of the electric field from the effective non-linear Maxwell equation
in the non-linear optical medium \cite{mcp}
\be
\nabla^2 \vec{E}^{(3)} - \nabla(\nabla \cdot \vec{E}^{(3)} ) + \frac{\omega_3^2}{c^2} \vec{\epsilon} \cdot \vec{E}^{(3)} = 
\frac{1}{c^2} \frac{\partial^2 \vec{\mathscr{P}}^{(3)}}{\partial t^2},
\ee
and thus, by neglecting the terms that are irrelevant over the length-scales on which the 3-rd wave changes are
long compared to the wavelength of the third wave, we deduce that the above schematic equation for the coupling of amplitudes $A_i$
becomes, more explicitly, the following three equations \cite{mcp}
\begin{subequations}
 \label{e:NLO}
\be
\frac{\rd A_i^{(3)}}{\rd z} = b_{ijk} A_i^{(1)} A_j^{(2)}, \quad \omega_3 = \omega_1 +\omega_2, \quad  k_3 = k_1 +k_2,
\ee
as well as
\be
\frac{\rd A_i^{(1)}}{\rd z} = b_{ijk} A_i^{(3)} A_j^{(2)*}, \quad \omega_1 = \omega_3 -\omega_2, \quad  k_1 = k_3 - k_2,
\ee
and finally
\be
\frac{\rd A_i^{(2)}}{\rd z} = b_{ijk} A_i^{(3)} A_j^{(1)*}, \quad \omega_2 = \omega_3 -\omega_1, \quad  k_2 = k_3 - k_1.
\ee
\end{subequations}
These are the crucial non-linear equations that we want to export for the case of an intrinsic triple interference in quantum gravity
(gravitization of quantum theory).

{Here we also briefly sketch the discussion of a triple interference experiment in this non-linear optics set-up, 
closely following  \cite{Namdar:2021czo}.}
It is explicitly stated in \cite{Namdar:2021czo} that in order to measure 
 $I_3$ (called the Sorkin Parameter in \cite{Namdar:2021czo}) three (or more) paths are sent through
the apparatus and all combinations of paths are blocked and unblocked using the path blockers. 
The authors of \cite{Namdar:2021czo} place a detector in a
path in order to estimate the probability to
detect a particle. Then from these measurements the authors of \cite{Namdar:2021czo}
compute the Sorkin Parameter $I_3$ from the defining expression 
$I_{3}(A,B,C) =P_{3}(A,B,C) -P_{2}(A,B)-P_{2}(B,C) -P_{2}(C,A)+P_{1}(A)+P_{1}(B)+P_{1}(C)$.
The authors of \cite{Namdar:2021czo} also discuss what they call a nonlinear triple slit in which a pump beam of wavelength $\lambda$
interacts with a signal and an idler beam, both with wavelengths $\lambda$ in a nonlinear crystal.
We refer the reader for the explicit experimental set-up used in \cite{Namdar:2021czo} to that reference, and in particular Figure~1 of \cite{Namdar:2021czo}.
The experimental results concerning the non-zero value of $I_3$ are presented in Figure~2 of \cite{Namdar:2021czo}.

\section{Triple interference: gravitizing the quantum}
\label{s:3nterference}
Triple interference in non-linear optics fundamentally stems from the fact that the electromagnetic excitations are traveling through an active and (in particular reactive) medium.  Analogous to this situation, we now present
a summary of an explicit model of gravitization of the quantum which leads to an intrinsic triple interference in a varying gravitational field, based on ~\cite{Berglund:2022skk}.

The fundamental insight is that in the context of gravitizaton of quantum theory, the expression for the probability is
generalized from a quadratic expression to a locally (in the space of states, not in spacetime!) quadratic expression,
which gravitizes the Born rule (in the manner the local spacetime Minkowski metric is gravitized to a general pseudo-Riemannian metric
in general relativity)\cite{Minic:2002pd, Minic:2003en, Minic:2004rj,  Jejjala:2007rn, Minic:2020zjb}.~\footnote{We note as an aside that the idea of state dependent operators, which we are introducing, have also been employed in a holographic context to resolve tension in black hole thermodynamics~\cite{Raju}.  Here state dependent boundary operators are able to probe otherwise inaccessible regions beyond the black hole horizon.}
Furthermore, in order to have such a generalized expression for the Born rule, one needs to
generalize the very kinematics of quantum gravity. In particular, we need multiple (observer dependent) Hilbert spaces that can interact with each other.

The generalized probability in this approach to quantum gravity is thus given by
\be
  P =    g_{ab}(\psi)\, \psi_a \psi_b \equiv \delta_{ab}\, \psi_a \psi_b + \gamma_{abc}\, \psi_a \psi_b \psi_c+\dots
   ,
\label{e:deformP}
\ee
where $a,b,c$ are state-space\footnote{In standard QM framework, the $\psi_a$ would be vectors in a Hilbert (linear vector) space; the explicitly included cubic term and the omitted higher order terms in~\eqref{e:deformP} invalidate this linearity and imply the state-space to be nontrivially curved,
and to have, in general, non-trivial topology.
By presuming $\gamma_{abc}$ to parametrize {\em\/small deformations\/} and omitting higher order terms, the $\psi_a$ represent vectors that are locally tangent to the true curved state-space.} indices. This is consistent with the general geometric formulation of
quantum theory discussed in the next section \cite{Minic:2003en, Minic:2004rj}. Since this expansion may be understood as an expansion in Riemann normal coordinates, the coefficients $\gamma_{abc}$ could be understood as specifying a connection/Christoffel symbol.
The fundamental equation that relates the non-linear (or more precisely, multi-linear) evolution of these state-spaces reads as the following {\it schematic} equation
(given the conservation of energy and momenta, and thus unitarity, in a generalized sense compatible with the
above generalized probability)
\be
\frac{\rd \psi_a}{\rd \tau} = \gamma_{abc}\, \psi_b \psi_c,
\label{e:deformDpsi}
\ee
where $\tau$ is the appropriate evolution parameter.
We note the similarity of these equations to those of Nambu quantum theory \cite{Minic:2002pd, Minic:2020zjb}, as well as the system~\eqref{e:dA=AA} describing nonlinear optics in nontrivial media.

To describe the intrinsic triple interference phenomenon and following the analysis of the nonlinear example~\eqref{e:NLO}, we introduce three distinct state-spaces, and also distinguish the corresponding state-functions by a parenthetical superscript.
 Using the deformaton~\eqref{e:deformP} and~\eqref{e:deformDpsi}, the coupled system of dynamical equations may then be written as:
\begin{subequations}
 \label{e:3slits}
\be
 \frac{\rd \psi_a^{(3)}}{\rd \tau}
 =\gamma_{abc}\,\psi_b^{(1)} \psi_c^{(2)},
\label{e:3=12}
\ee
accompanied by (including the canonical linear term)
\be
\frac{\rd \psi_a^{(1)}}{\rd \tau}
 = \delta_{ab}\,\psi_c^{(2)*}
  + \gamma_{abc}\,\psi_b^{(3)} \psi_c^{(2)*},
\label{e:1=23}
\ee
and finally (also including the canonical linear term)
\be
\frac{\rd \psi_a^{(2)}}{\rd \tau}
 = -\delta_{ab}\,\psi_c^{(1)*}
  - \gamma_{abc}\,\psi_b^{(3)} \psi_c^{(1)*}.
\label{e:2=13}
\ee
\end{subequations}
The subscripts $a,b,c$ here label states within the state-space indicated by one of the superscripts, $(1), (2)$ or $(3)$; we proceed assuming that the three state-spaces are isomorphic and so have identically labeled states. (The signs are consistent with real $\psi$'s and the canonical Schr\"{o}dinger equation in the
case of zero $\gamma_{abc}$.)
This analysis may be adopted also to discuss real state-spaces, e.g., the real and imaginary parts of the complex state-spaces (if such a separation is possible) as
used in the canonical case, which should be recovered in the limit $\psi^{(3)}_a\to1$ and $\gamma_{abc}\to0$.

Crucially $\psi^{(3)}$ is one in the usual non-gravitational case.
So, it could be an exponent of an (imaginary) unit times the Planck length over the characteristic scales
of interference, for example. Similarly for the ``cubic'' tensor $\gamma$ --- it should be proportional
to the Planck length over the characteristic length of interference. The cubic tensor characterizes
the gravitized case, in which intrinsic triple interference is expected. 

Note that we are proposing a fundamental change in the kinematics
of quantum gravity. Instead of one Hilbert space, with the rigid
canonical quantum geometry, linearity, unitarity and the canonical Born rule, we advocate for observer-dependent Hilbert spaces, and
thus observer-dependent spacetimes, all parts of a quantum spacetime with a dynamical geometry, and thus a dynamical quantum geometry. In essence, we are proposing quantum relativity and quantum diffeomorphism invariance, and a higher categorical description of the kinematics of quantum gravity, viewed as ``gravitization of the geometry of quantum theory''. In the context without quantum gravitational degrees of freedom, this structure reduces to the rigid structure of canonical quantum theory. All this is consistent with 
previous research on metastring theory~\cite{Freidel:2015uug, Freidel:2016pls, Freidel:2013zga, Freidel:2014qna, Freidel:2015pka, Freidel:2017xsi,  Freidel:2018apz,  Freidel:2019jor, Minic:2020oho,  Freidel:2017wst, Freidel:2017nhg} as well as work in relative locality.

One thing is evident: the moment we go beyond the usual vector space of quantum theory and quantum field theory
(including quantum field theory in curved spacetime) as well as pretty much all major approaches to quantum 
gravity (canonical/loop, canonical string theory, semiclassical quantum gravity, Euclidean quantum gravity, etc),
we are changing the very kinematics of quantum gravity.
If triple, and higher order, interference in a gravitational fiels is non-zero (and that is the crucial empirical question)
then one needs to use the new concept of ``dynamical Hilbert spaces'' or a higher category of vector spaces (2-vector spaces, for example).

Also, the observables are changed: instead of having only transition amplitudes (including the S-matrix, properly defined in a general curved backround)
associated with the quadratic nature of the Born rule in quantum theory, in the context of gravitized quantum theory (quantum gravity)
one needs triple and higher order generalizations of the S-matrix and transition amplitudes.
In principle, there is an infinite number of such new quantities (which, given the non-linear nature
of gravity, are likely to resum into a fundamental non-linear expression, reminiscent of what happens
in the context of Einstein's general theory of relativity).

As a conclusion to this section we sketch a proposal for a triple interference experiment in a varying gravitational field
that is inspired by the triple-path interference experiment in non-linear optics \cite{Namdar:2021czo}.
(The gravitational field plays the role analogous to a non-linear optical medium.)
Thus we propose to measure 
 $I_3$ by having three (or more) paths sent through
the apparatus in a varying gravitational field 
where all combinations of paths are blocked and unblocked using the path blockers. 
We envision placing a detector in a
path in order to estimate the probability to
detect a particle in a gravitational field. Then from these measurements we should
compute the Sorkin Parameter $I_3$ from the defining expression 
$I_{3}(A,B,C) =P_{3}(A,B,C) -P_{2}(A,B)-P_{2}(B,C) -P_{2}(C,A)+P_{1}(A)+P_{1}(B)+P_{1}(C)$.

The crucial experimental question is : Is $I_3 \neq 0$?

{ In section 8 we will analyze the model proposed in the preceding discussion and compare it to the discussion of
triple interference found in the context of non-linear optics.
Note that the non-linear optics paper \cite{Namdar:2021czo} explicitly states that their effective non-linear 
equation was solved numerically. 
We will proceed in a more analytic fashion in section 8.
This is possible because  Jacobi elliptic functions~\cite{WW} satisfy very similar non-linear equations to our proposed model of triple interference in the presence of gravity:
\be
\frac{\rd\,\mathrm{sn}(z)}{\rd z} = \mathrm{cn}(z)\,\mathrm{dn}(z), \quad
\frac{\rd\,\mathrm{cn}(z)}{\rd z} = -\mathrm{sn}(z)\,\mathrm{dn}(z), \quad 
\frac{\rd\,\mathrm{dn}(z)}{\rd z} = - k^2 \mathrm{sn}(z)\,\mathrm{cn}(z)
\label{e:Jacobi}
\ee
where $k$ is the modulus. These are also familiar from Nambu quantum theory \cite{Minic:2002pd, Minic:2020zjb}. }
Now we proceed to address some possible concerns regarding the proposed generalization of the Born rule.

\section{There be dragons: the robustness of the Born rule}
\label{s:dragons}
Modifying the Born probability rule \cite{bornrule} is fraught with complication, which is somewhat surprising as the Born rule was originally an ad hoc postulate of quantum mechanics (for a review, see \cite{earman}).  There was no derivation of the Born rule from other principles --- it was just a rule that worked experimentally.  Hence during the twentieth century numerous physicists explored modifications to the rule.  However, despite being at first innocuous, these modifications led to violations of causality or unitary.  And unlike the Born rule, causality and unitarity are very well grounded, extremely well-tested principles of physics --- violating them generally makes a theory considered sick.  Modifying the Born rule has therefore been approached with a well-founded skepticism over the past few decades, to the degree that quantum mechanics has been called an ``island in theory space''
\cite{island}.  

The surprising resistance of the Born rule to modification gave rise, starting with the work of Gleason \cite{gleason}, to a parallel track of theoretical exploration that attempts to derive the Born rule from more basic assumptions, instead of leaving it as an ad hoc addition to quantum mechanics.  Since these works stand as mathematical theorems, any adjustment to the Born rule must then violate one of these more fundamental assumptions.  It is therefore useful to understand how quantum gravity may modify these assumptions, providing a natural starting point for a theoretical investigation of the Born rule from the viewpoint of quantum foundations.  Since there are a number of derivations from different assumptions, an exhaustive review is beyond the scope of this paper (although we fully acknowledge that such an exhaustive examination must eventually be done). Instead, we will simply give an illustrative example, focusing on the assumptions from one such recent derivation, the work of Masanes, Galley, and Muller \cite{mgm}. This work derives the Born rule from three postulates:
\begin{enumerate}
\item The set of states of every physical system can be described by a complex, separable Hilbert space $\mathcal{C}^d$.  Pure states are the rays, i.e. $\psi \in P \mathcal{C}^d$.
\item The reversible transformations for pure
states are the unitary transformations $\psi \rightarrow U \psi$ with $U \in U(d)$.
\item If one has a composite system built of two subsystems $A,B$, with state spaces $\mathcal{C}^a$ and $\mathcal{C}^b$ respectively, the pure states of the joint system are the rays of the tensor product space $\mathcal{C}^{ab}=\mathcal{C}^a \otimes \mathcal{C}^b$.
\end{enumerate}

The first postulate and second postulates deal with the structure of the state space of individual quantum systems and its evolution.  Since we are inherently concerned with composite systems, we leave these postulates unmodified.  The third postulate however, is immediately problematic when considering quantum gravity.

The third postulate is an assumption about factorizability of a joint Hilbert space.  Factorizability can be re-expressed generally as the ability to find two commuting sub-algebras of operators that represent observables in the joint Hilbert space.  We can then use those sub-algebras to factorize the Hilbert space into a tensor product structure.  Factorizations are not unique --- there can be many different factorizations of the same joint Hilbert space.  The most common factorization most physicists are familiar with is that of local quantum field theory.  We have the notion of local field operators $\hat{\phi}(x)$ which satisfy microcausality: on a spacelike surface $[\hat{\phi}(x), \hat{\phi}(y)]=0$.  Microcausality, which in turn encapsulates the idea of cluster decomposition~\cite{weinbergQFT}, implies that the local Hilbert spaces are independent and the total Hilbert space factorizes into the usual tensor product.    

The above construction is immediately problematic when considering gauge theories with associated constraints.  In a gauge theory such as electromagnetism, the gauge invariant field operators are non-local and hence there can be non-zero commutators between gauge invariant operators at spacelike separation.  This obstruction can be removed by introducing edge modes for bounded regions, which restores the factorizability macroscopically~\cite{Donnelly:2016auv}.  Since gravity can be represented as a theory with a gauged Poincare algebra, the same approach can be used for a gravitational theory of a metric on a background spacetime.  

In a full quantum gravity theory, however, the above construction is not enough, as topology change, inherent in any quantum gravity theory, is not accounted for in an edge mode type construction.  When the definition of the boundary of a region is part of the dynamical construction, geometric edge modes on a fixed boundary are not enough to ensure factorizability.  A wormhole connecting two otherwise disjoint regions of spacetime would, for example, can break the edge mode construction and yield a non-zero commutator between naively spacelike separated field operators.  A similar result has been found in the context of AdS/CFT, where the field operators in two asymptotically AdS regions that are otherwise disjoint acquire a non-zero commutator upon the introduction of wormholes connecting the regions through the bulk~\cite{Marolf:2020xie}.

Independent of the precise mathematical construction, any approach which trivially introduces non-zero commutators between spacelike separated operators in a standard d-dimensional spacetime suffers an immediate and glaring problem: it violates causality and leads to superluminal signaling and closed causal loops.  Any quantum gravity theory that satisfies the chronology protection conjecture (which we will assume) would forbid such constructions.  We cannot therefore modify the third postulate and hence the Born rule by considering only modifications of operators on a spacetime with standard microcausality and locality.  Since quantum gravity however is a quantum theory of spacetime, there would seem to be no out --- quantum gravity respects the Born rule.  

There is, however, an out if the quantum gravity theory in question also modifies the structure of spacetime itself in such a way that the local $\mathcal{R}^n$ structure inherent in general relativity, which can be described only by usual commuting coordinates $x$, is extended.  Such a structure occurs naturally in metastring theory, where the underlying spacetime is not the usual spacetime of quantized general relativity, but is instead a modular spacetime described by a Born geometry,as found in metastring theory~~\cite{Freidel:2015uug, Freidel:2016pls, Freidel:2013zga, Freidel:2014qna, Freidel:2015pka, Freidel:2017xsi,  Freidel:2018apz,  Freidel:2019jor, Minic:2020oho,  Freidel:2017wst, Freidel:2017nhg}. Born geometry requires two distinct sets of coordinates to  describe, $x$ and $\tilde{x}$.  In such a scenario, one can establish that commutators of local operators e.g. $[x,y]$ commute, preserving our familiar spacetime causality, but commutators of the form $[x,\tilde{x}]$ do not, which allows for a violation of factorizability and a modification of the Born rule.  Indeed, as indicated by the arguments above such a construction may not only be sufficient but actually necessary for any modification to the Born rule that is not immediately ruled out. 
 
We now turn to a brief summary of metastring theory and Born geometry.
We emphasize the need for a kinematical generalization of the canonical quantum geometry in the context of quantum gravity.
Only in that context does the Born rule find its generalization, thus avoiding its apparent robustness found in the context of canonical quantum theory.

\section{Gravitizing the quantum is Quantizing gravity}
\label{s:GQ=QG}

\subsection{Top-down view from string theory}
Modified quantum evolution is a typical approach in quantum gravity phenomenology, and indeed the entire area of searching for spacetime symmetry violation can be reframed as looking for quantum gravity induced modifications of propagators~\cite{Addazi:2021xuf}.  What is new  in our approach is the intrinsic non-linearity/multi-linearity of the kinematics of quantum theory which has not been significantly explored.  However, one can construct arguments directly from string theory supporting non-linear modifications.  In that case the non-linear cubic theory is string field theory: For open strings, this is the Witten cubic open string field theory~\cite{Witten:1985cc}; for closed strings there is a non-associative cubic string field theory proposed by Strominger~\cite{Strominger:1987bn} and a more canonical non-polynomial closed string field theory of Zwiebach~\cite{Zwiebach:1992ie}; for more recent references, c.f.~\cite{SFT1,SFT2,SFT3,SFT4}. String theory though has state operator correspondence, so from the viewpoint of \cite{Berglund:2022skk},
the famous ``pants'' diagram of string theory (the cubic interaction vertex of string field theory) 
\cite{Polchinski:1998rq} can be viewed as an example of a non-linear modification, or gravitization, of quantum mechanics.

Indeed, as we have explained in our previous paper~\cite{Berglund:2022skk}, the cubic vertex in string field theory (SFT) can be interpreted as intrinsic triple interference only on the branching underlying {\em worldsheet} itself. There is no violation of the Born rule in the background, {\em target} spacetime. Similarly, the cubic vertex in ordinary quantum field theory (QFT), can be interpreted as intrinsic triple interference only from the point of view of the branching underlying worldline. In the background classical (target) spacetime, the Born rule is still valid, unchanged. This is one of the essential insights that we have: both the canonical QFT and SFT can be viewed, from the viewpoint of the underlying worldline and worldsheet, respectively, as quantum gravity theories in, either $0{+}1$ or $1{+}1$ dimensions. Only from the point of view of the branching (topology changing) worldline or worldsheet, respectively, can one talk about intrinsic triple interference, by interpreting the classical 
QFT of SFT equations of motion, with a cubic vertex, as the Wheeler-DeWitt equations (Hamiltonian constraints in the context of either $0{+}1$- or $1{+}1$-dimensional quantum gravity) on the worldline and worldsheet, respectively. Thus, it is natural to expect intrinsic triple (and, given the nonlinear nature of gravity) higher order intrinsic interference in quantum gravity in $3{+}1$ dimensions. That is why we expect quantum gravity to be a ``gravitization of quantum theory'', because only in that context can one realize intrinsic triple and higher order interference.

As already indicated in the previous section, our proposal can be made precise in the context of {\em\/metastring\/} theory~\cite{Freidel:2015uug, Freidel:2016pls, Freidel:2013zga, Freidel:2014qna, Freidel:2015pka, Freidel:2017xsi,  Freidel:2018apz,  Freidel:2019jor, Minic:2020oho,  Freidel:2017wst, Freidel:2017nhg, Freidel:2021wpl} (which includes a useful phenomenological limit, called the {\em\/metaparticle\/}~\cite{Freidel:2018apz}) that captures intrinsic non-commutative 
(as well as non-associative~\cite{Gunaydin:2013nqa}) features
of string theory, leading to many novel and interesting predictions~\cite{Berglund:2022qcc}. This theory can be understood as defining a precise notion of 
the {gravitization} of quantum geometry, consistent with unitarity and causality,
in the sense that quantum mechanical structures that are traditionally fixed become dynamical.
Note that the modified evolution equations discussed in this section can be made more precise in the non-associative cubic string field theory 
\cite{Strominger:1987bn},
as well as in metastring theory~\cite{Minic:2020oho, Berglund:2020qcu}.  In fact, intrinsic triple-path interference can be directly related to the non-associativity 
\cite{Barnum:2014ysa, tripleint} which occurs in metastring theory and cubic closed string field theory, lending further credence to our phenomenological proposal.

For example, in the context of one proposal for a non-perturbative formulation of metastring theory one encounters the
following matrix model \cite{Berglund:2020qcu, Minic:2020oho} (to be understood as a metastring-inspired generalization
of the matrix formulation of canonical quantum theory)
\begin{equation}
 \S_{\text{ncM}}\,{=}\,\frac{1}{4\pi}  
\int_{\tau} 
\big(\pa_{\tau} \X^i [{\X}^{j},  {\X}^{k}] g_{ijk} - [{\X}^{i},  {\X}^{j}][{\X}^{k},  {\X}^{l}] h_{ijkl}\big),
\end{equation}
with 27 bosonic $\X(\tau)$ matrices (with supersymmetry emerging in 11 dimensions) and with $\tau$ denoting a ``world-line`` parameter.
(These requirements are imposed by the general structure of metastring theory~\cite{Freidel:2015uug, Freidel:2016pls, Freidel:2013zga, Freidel:2014qna, Freidel:2015pka, Freidel:2017xsi,  Freidel:2018apz,  Freidel:2019jor, Minic:2020oho,  Freidel:2017wst, Freidel:2017nhg, Freidel:2021wpl}, in which apart from spacetime coordinates $x$ one has
dual spacetime coordinates $\tilde{x}$ which do not commute with $x$. The pair $x$ and $\tilde{x}$ is put together into a doublet $\X$.)
The relevant backgrounds $g_{ijk}(\X)$ and $h_{ijkl}(\X) $ (which, in general depend on $\X$) should be determined by the matrix
renormalization group
(RG) equations (in analogy with the condition of conformal invariance in perturbative string theory, that leads to the Einstein equation for 
the metric background).
These backgrounds $g_{ijk}(\X)$ and $h_{ijkl}(\X) $ are captured by Born geometry, which unifies the symplectic, bi-orthogonal and doubly-metrical backgrounds
of the metastring~\cite{Freidel:2015uug, Freidel:2016pls, Freidel:2013zga, Freidel:2014qna, Freidel:2015pka, Freidel:2017xsi,  Freidel:2018apz,  Freidel:2019jor, Minic:2020oho,  Freidel:2017wst, Freidel:2017nhg, Freidel:2021wpl}, and which is responsible for its intrinsically non-commutative and phase-space-like
nature.
In particular, the doubly-metrical background contains the usual Einstein metric of general relativity. 
This formulation fits the new concept of ``gravitization of quantum theory'' as well as the idea that dynamical Hilbert spaces or 2-Hilbert spaces (here represented by matrices) are fundamentally needed
in quantum gravity \cite{twohilbert}.
In particular, the matrices $\X$ could be understood as collections of vectors and
as such they can be understood to be elements of such 2-Hilbert spaces (the category whose elements belong to
the category of Hilbert spaces). This would be an intrinsically dynamical formulation of interacting Hilbert spaces with
a direct connection to a theory of quantum gravity (string theory in its metastring formulation). Such a formulation is
naturally related to a general geometric formulation of quantum theory, which in turn is directly related to the Born rule and its
generalization and the above 
discussion of multi-path interference.  We now turn to this geometric viewpoint.

\subsection{Bottom-up view from geometric quantum mechanics}
The geometry of canonical quantum theory is very rigid and is represented by the complex projective geometry~\cite{Minic:2003en}.
Geometrically, the Born rule characterizes odd-dimensional spheres (implied by the Fisher metric in the space of distinguishable events), which in turn can be viewed as $U(1)$ bundles over complex projective spaces~\cite{Minic:2004rj}. The metric structure of the complex projective space so-specified by the Born rule
and its complex structure are compatible with the symplectic structure
 (which in a singular limit reduces to the classical symplectic structure)~\cite{Minic:2003en, Minic:2004rj}. In particular, given the Born rule this is the well-known Fubini-Study metric, which is an Einstein metric and satisfies the vacuum Einstein equations with a dimension-dependent cosmological constant.  Notably, the vacuum Einstein equations are expressly not state (position) dependent, to wit, there is no local state (position) dependent source. 

Following the work of~\cite{Minic:2003en} (and the suggestion of Ashtekar and Schilling~\cite{Ashtekar:1997ud}), 
one can ``gravitize'' this rigid geometry of quantum theory by making the Fubini-Study metric dynamical~\cite{Minic:2003en, Minic:2004rj, Jejjala:2007rn}.  
Since in the absence of state dependence the vacuum Einstein equations are satisfied, then one natural candidate for the governing dynamical equation for the Fubini-Study metric with local state dependence is simply the full quantum Einstein equation~\cite{Berglund:2022skk}:
\be
G_{ab}[g_{cd}^{FS}]+ \Lambda\,g^{FS}_{ab}=G_{QG}\,T_{ab}(\psi)
\ee
where $T_{ab}(\psi)$ is a local function on the space of states and $G_{QG}$ is some dimensionful coupling that would vanish as we return to the Born rule. 
(These Einstein-like equations are implied by the general matrix theory discussed above. The geodesic equation that follows from this
Einstein-like equation defines the non-linear evolution of Hilbert spaces stated in section~\ref{s:3nterference}, with $\gamma_{abc}$ playing the role of the connection
symbol. It was proposed in \cite{Minic:2003en, Minic:2004rj} that generalized non-linear Grassmannians play the role of complex projective
spaces in this context. We will comment on the generalized geometry of gravitized quantum theory in what follows.) $G_{QG}$ is technically independent of but presumably related to $G_N$, Newton's constant. While the classical and quantum theories are different (and operate in different spaces), we may consider flat-space quantum field theory with the Born rule as our unperturbed, or ``un-gravitized'' theory. Then, the vanishing of all geometrical and topological quantum gravity effects corresponds to the {\em\/double limit\/}  $G_N,G_{QG}\,{\to}\,0$, which indicates a relation between the two.   In summary, the modification to the Fubini-Study metric is in complete analogy with the transition from the
fixed Minkowski geometry of special relativity (a solution of the vacuum Einstein equations) to the dynamical geometry of general relativity where there is now a local source term.

{ Here we offer a visualization of the gravitized quantum geometry: The usual quantum geometry is that of a complex projective space.
Let us take the example of the one dimensional complex projective space --- the Bloch sphere. The gravitized geometry would involve a 
generalization of this sphere to a Riemann surface with an arbitrary number of handles (genus). This ``foamy'' structure (in the sense of Wheeler, albeit
in the space of quantum states and NOT in spacetime) is our proposal for the gravitized quantum geometry. The handles could be associated with extra 
observer dependent Hilbert spaces
which now interact with the canonical Hilbert space. The Born rule is generalized so that higher order inerference is allowed.
These observer dependent Hilbert spaces will be represented later in this paper in a toy model that resembles Nambu quantum
theory~\cite{Minic:2002pd, Minic:2020zjb}. The main point is that in quantum gravity one works with observer dependent Hilbert spaces, and that the kinematical category that is
associated with the very description of quantum state in quantum gravity is generalized, in analogy what happens when one goes from
special theory of relativity to general theory of relativity.}

As a final comment on the general geometry of quantum theory, we recall that the quantum clock embodies the relation between the Fubini-Study metric of the complex projective spaces and the
time interval multiplied by the uncertainty in energy (see~\cite{Minic:2003en, Minic:2004rj, Jejjala:2007rn} and the original work of Aharonov and Anandan~\cite{aa}).
The basic equation here follows from unitary evolution and reads as
$
2 \hbar\, \rd s_{FS} = \Delta E\, \rd t, 
$
where $\rd s_{FS}$ characterizes the Fubini-Study metric of complex projective
spaces (which is in turn directly related to the Born rule and the Fisher metric in the space of quantum states\cite{Minic:2003en}),  $\Delta E$ is the dispersion of energy, and $\rd t$ the time differential.
In principle, in the context of a gravitized quantum theory, the Hamiltonian becomes effectively state dependent $H \to H_{\psi}$ and so the 
energy dispersion becomes
state dependent $ \Delta E  \to (\Delta E)_{\psi}$ and thus the quantum geometry becomes dynamical $\rd s_{FS} \to \rd s_{FS} (\psi)$, and 
subject
to the above quantum Einstein equation.
(The intuition here is as follows: the clocks are
dynamically influenced by gravity in general relativity, but the clocks are fundamentally quantum objects, as indicated by the above equation that relates the geometry of canonical quantum theory to the measurement of time. Thus, in the full quantum gravity context, dynamical quantum clocks call for a dynamical quantum geometry.)
What would be an experimental probe of this precise quantum statement regarding the quantum geometry?
Here we recall the recently developed optical lattice atomic clocks which have been used to measure the gravitational red-shift to a remarkable accuracy 
\cite{katori}. In this case one tests corrections to the red-shift formula $\frac{\Delta \nu}{\nu} = (1 + \alpha) \frac{\Delta U}{c^2}$,
where $\nu$ is the frequency and $U$ the gravitational potential, and the correction factor $\alpha$ (equal to zero in general relativity)
is bounded to the order of $10^{-5}$.
In view of the relation between quantum geometry and quantum clocks, the ultra-precise experimental tests of the validity of general relativity can be used 
in principle to constrain the gravitization of quantum geometry.

\section{The scale of triple interference involving gravity}
\label{s:scale}
The usual expectations on quantum gravity tie any putative quantum gravity effects to the Planck scale $l_P = 10^{-35}m$ (and Planck
energy $E_P \sim 10^{19} GeV$)
which is experimentally prohibitive, to say the least.  While this is still possible, 
in this section we present an argument for the scale of triple interference to perhaps be of the order of the Higgs scale $10^{-19}m$ which is far more experimentally accessible.

First, we note the recent progress in understanding of the vacuum energy problem in quantum field theory and
quantum gravity. We will not review in detail the arguments presented in \cite{Freidel:2022ryr}, \cite{Berglund:2022qsb}.
The most important point for us is that the two scales that appear in classical Einstein's equations, the cosmological, Hubble
scale, $l$ (the observed cosmological constant $\lambda_{cc}$) and the scale associated with the Newton gravitational constant, the Planck scale $l_P$ 
appear together in the formula for the vacuum energy scale $\rho_0$ (which determines the cosmological constant $\lambda_{cc} \sim \rho_0 l_P^2$).
What was derived in \cite{Freidel:2022ryr} is that energy scale associated with $\lambda_{cc}$, denoted by $l_{\lambda}$
\be
l_{\lambda} \sim \sqrt{l\, l_P},
\ee
which corresponds to what is observed.
Now, the reasoning that leads to this formula can be repeated for another energy scale in the matter sector which determines the
vacuum --- the Higgs energy scale $E_H$ --- which also turns out to be related to the geometric mean of the $E_{\lambda}$ and $E_P$ energy scales
\be
E_H \sim \sqrt{E_{\lambda} E_P}.
\ee
We want to claim that this is one of the relevant scale for the effect of triple interference in quantum gravity.

One argument is statistical: The derivation of the above relation for the scale $l_{\lambda}$ is based on the
understanding of the size of the momentum space in 4d, $\epsilon^4$, based on the quantization of the phase space
$l^4 \epsilon^4 \sim N$, where $N$ the number of degrees of freedom, is given by the Bekenstein bound (gravitational holography),
which asserts, for the case of the cosmological horizon, that, $N \sim l^2/{l_{P}}^2$. Then the vacuum energy 
$\rho_0 \sim \epsilon^4 \sim N/l^4 \sim 1/(l^2 l_P^2)$
from which the above formula $l_{\lambda}\sim 1/{\epsilon} \sim \sqrt{l\, l_P}$ follows. (There is a precise calculation associated with this simple scaling argument~\cite{Freidel:2022ryr}, \cite{Berglund:2022qsb}.) The derivation of the formula for $E_H$ proceeds along similar lines.
(Note, $l \sim 10^{27}m$, $l_{\lambda} \sim 10^{-4} m$, and $E_{\lambda} \sim 10^{-3} eV$ and $E_H \sim 1 TeV$.)
Note that the order of magnitude of $N \sim 10^{124}$ in the observed Universe. 

Be that as it may, note that in each direction of spacetime one has in 4d, $N^{1/4}$ ``units'', that scale as $N_i \sim 10^{31}$ ($i=1,2,3,4$), which is
not that far from the Avogadro number associated with matter degrees of freedom. Note that for a statistical system
made of $N$ independent units, fluctuations scale as $1/{\sqrt{N}}$, so $N$
has to be large for the reasons of stability of the overall system, and also the square root of the dispersion scales as $\sqrt{N}$,
and thus the scales that are off by a factor of $\sqrt{N_i}$ from the Planck scales are natural by this argument.
Thus the Higgs scale, which is $10^{16}$ away from the Planck scale ($\sqrt{N_i} \sim 10^{15.5}$)

Another argument that leads to the same conclusion runs as follows.  As pointed out in the previous section the gravitization of the quantum suggests that the geometry of complex projective spaces is generalized, so that
the unitary geometry is replaced by quantum diffeormorphisms.
In principle, one has to allow for topology change, and thus one could visualize such spaces as,
in the simplest case of the one dimensional complex projective space, also known as the Bloch sphere, a sphere
with an arbitrary number of handles.
The handles can be associated with other Hilbert spaces needed for the mathematical description of gravitization of quantum geometry,
as presented in a toy model in the next section. 

The important point here is as follows: it is well known in statistical physics and in quantum field theory how to sum
over contributions of closed diagrams. One simply gets an exponent of the partition function associated with a closed loop
(a closed handle in our case). This follows from simple combinatorics.
For example the vacuum partition function in field theory $Z_v = \exp(Z_{S^1})$, where $S^1$ is the
circle associated with a vacuum loop traced by a particle. The same applies in string theory (just replace $S^1$ by a torus $T^2$) --- this
type of formula is very important for the argument made in  \cite{Freidel:2022ryr}.

In our case one has a ``Wheeler-like foam" in the space of quantum geometry, and given the above formula,
the usual path integral measure $e^{iS}$ should be replaced after summing over handles by 
$e^{e^{iS}}$, where in the approximation of a dilute gas of handles we have taken that the
effective partition function is just the canonical one.
Thus, if we sum over handles in the foamy quntum space, from the point of view of the
canonical complex geometry we have an effective action which is essentially $e^{iS}$.
In the euclidean language that means that the effective action at some scale sensitive to gravity can be
{\it exponentially removed} from the natural scale of Planck gravity.
Thus, the Higgs scale can be the place where one should see effects of quantum gravity, given by
the approach based on gravitization of quantum geometry.
(Please note that, essentially, we are claiming the hierarchy of scales between the Higgs and
the Planck scale is a quantum gravity effect, associated with gravitization of the quantum.)\footnote{Another comment might be
useful here: in this approximation of the gravitized quantum theory, the effective Feynman-like measure involves an
exponent of an exponent. Thus the number of degrees of freedom should go as the exponential of an exponent of some
effective action. This seems to match the expression needed to explain the fine-tuning of the initial state in quantum cosmology,
as repeatedly emphasized by Penrose, which should match the $10^{10^{120}}$ degrees freedom in the observed universe.
The fine-tuning of the initial state involves picking one out of this many states. This seems to be achievable in the gravitized quantum proposal.}

Both of these arguments lead to the same place --- the Higgs scale.
Thus we need to allow for the possibility that triple interference may be observed in the context of gravity
at the Higgs scale. (Note that the TeV scale features prominently in the extra dimension proposals \cite{extradim}, where the
extra dimensions are envisioned, in the context of effective field theory, to address the hierarchy of scales between the Higgs scale
and the Planck scale. In our case, it is gravitization of the quantum that addresses that exponential hierarchy (combined with the
recent arguments about the cosmological constant problem \cite{Freidel:2022ryr} and also, the gauge hierarchy problem \cite{Berglund:2022qsb}.)

\section{Phenomenology: Modified transition amplitudes}
\label{s:toy}
We now turn to making these general considerations and ideas more experimentally concrete in terms of possible phenomenology and experimental tests.  Our first possible phenomenological signature is that of a modification to transition probabilities, which we illustrate with a toy model.  Let us consider a much simplified version of the system~\eqref{e:3slits}, where all three state-spaces are collapsed to a single state-function each, so we can drop the $a,b,c$ indices. Consider $\j^{(1)}+i\j^{(2)}$ forming a typically complex state-function of a standard quantum (test) object, with a ``kinematic'' bilinear term, $\w_0\,\j^{(1)}\,\j^{(2)}$, in the Hamiltonian.
 In turn, $\j^{(3)}$ may be assumed to be real, modeling an interactive (perhaps gravitational) background:
\begin{subequations}
 \label{e:3slitss}
\begin{alignat}9
\frac{\rd \j^{(1)}}{\rd \tau}
 &= \w_0\j^{(2)} +\g\,\j^{(2)} \j^{(3)}&
 &= \g\,\j^{(2)}\big(\tfrac{\w_0}\g +\j^{(3)}\big),
\label{e:1=23s}\\
\frac{\rd \j^{(2)}}{\rd \tau} 
 &= -\w_0\j^{(1)} -\g\,\j^{(3)} \j^{(1)}&
 &= -\g\,\j^{(1)}\big(\tfrac{\w_0}\g +\j^{(3)}\big).
\label{e:2=13s}\\
\frac{\rd \j^{(3)}}{\rd \tau}
  &=\g\,\j^{(1)} \j^{(2)},
\label{e:3=12s}
\end{alignat}
\end{subequations}
Note that $\j^{(1)}\eqref{e:1=23s}+\j^{(2)}\eqref{e:2=13s}$ implies
\begin{equation}
  0=\j^{(1)}\frac{\rd \j^{(1)}}{\rd \tau}+  \j^{(2)}\frac{\rd \j^{(2)}}{\rd \tau}
   =\frac{\rd}{\rd \tau}\big((\j^{(1)})^2+(\j^{(2)})^2\big)
   =\frac{\rd}{\rd \tau}\big|\j^{(1)}+i\j^{(2)}\big|^2,
\end{equation}
so that the (separate, standard) probability $|\j^{(1)}+i\j^{(2)}|^2$ is conserved in time.
The factorization in~\eqref{e:1=23s}--\eqref{e:2=13s} suggests using
 $\widetilde\j^{(3)}\!:=\!(\w_0/\g + \j^{(3)})$, which highlights the correspondence and (sign-twisted) analogy of this system~\eqref{e:3slitss} to the nonlinear optics example~\eqref{e:NLO}, or to the classical mechanics of a spinning top. Furthermore, comparison with~\eqref{e:Jacobi} implies that the simplified system~\eqref{e:3slitss} admits exact solutions in terms of Jacobi elliptic functions. Indeed:
\begin{subequations}
 \label{e:psi123}
\begin{alignat}9
 \j^{(1)}(t)
 &= \pm \tfrac{k\,\w}\g\;\mathrm{sn}(\w\,t\,,-k^2),&\quad
     \w^2&=\w_0^{~2}{+}2\g c_2, \label{e:psi1}\\
 \j^{(2)}(t)
 &= \pm \tfrac{k\,\w}\g\;\mathrm{cn}(\w\,t\,,\,-k^2),&\quad
     k^2&=\tfrac{2\g^2 c_1}{\w_0^{~2}{+}2\g c_2}, \label{e:psi2}\\
 \j^{(3)}(t)
 &= -\tfrac{\w_0}\g \pm \tfrac\w\g\,\mathrm{dn}(\w\,t\,,\,-k^2),&\quad
     \widetilde\j^{(3)}&=\pm \tfrac\w\g\,\mathrm{dn}(\w\,t\,,\,-k^2), \label{e:psi3}
\end{alignat}
\end{subequations}
where the four distinct branches of the exact general solution always have an even number of negative signs among the three indicated $\pm$ sign-choices. The integration constants $c_1,c_2$ are free to be determined by boundary conditions; a third integration constant occurs only in the combination $(t{-}t_0)$, and we set $t_0{=}0$ for convenience. It is worth noticing that a substitution of~\eqref{e:psi3} in~\eqref{e:1=23s} and~\eqref{e:2=13s} simplifies these equations so that $\g$ appears only implicitly, within the arguments of these functions.

\begin{figure}[htb]
 \begin{center}
 \includegraphics[width=160mm]{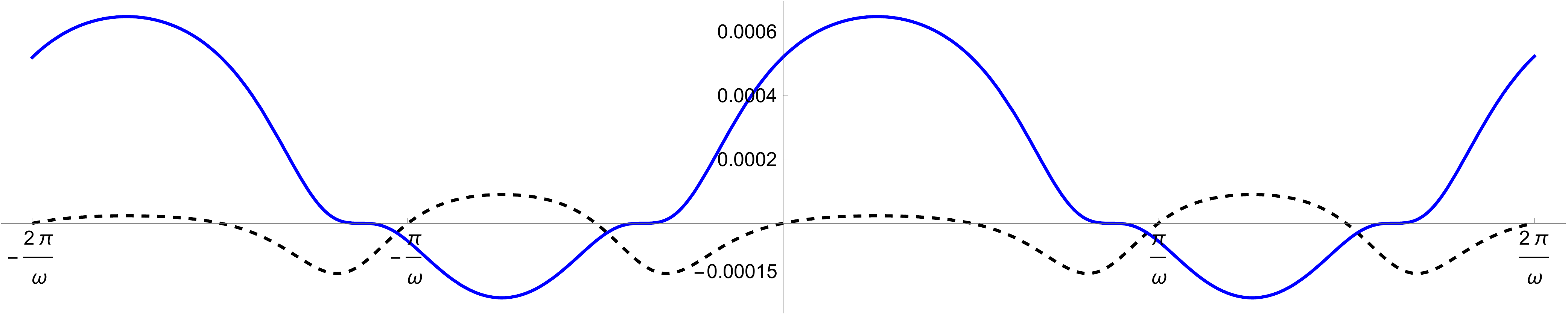}
 \end{center}
 \caption{The normalized triple interference measures $\kappa(1,2,3)$ (blue, solid) and $\widehat\kappa(1,2,3)$ (black, dashed) for $\w_0,c_1\,{=}\,1$, $c_2\,{=}.75$ and $\g\,{=}\,10^{-3}$,
 so $\w\,{\approx}\,1.000\,75$ and $k\,{\approx}\,\sqrt2\,\g$.}
 \label{f:kappa}
\end{figure}

Much as the canonical quadratic term in~\eqref{e:deformP} defines the standard QM probability
\be
 P=\d_{ab}\,\j_a\j_b~~\mapsto\quad
 P_2(i,j):=|\j^{(i)}+\j^{(j)}|^2,
\ee
by superposition ($\j_a\to\j^{(i)}+\j^{(j)}$), the cubical deformation~\eqref{e:deformP} defines for the simplified system~\eqref{e:3slitss}:
\be
 P_3(1,2,3):=\big|\j^{(1)}+\j^{(2)}+\j^{(3)}\big|^2
 +\Re\big[\g\,(\j^{(1)}+\j^{(2)}+\j^{(3)})^3\big],
\ee
where to lowest order in $\g$ the argument of $\Re$ uses ordinary superposition, uncorrected by further powers of $\g$. The quadratic term cancels in the triple interference quantity~\eqref{e:I3}, and we remain with  $I_3(1,2,3)=\Re\big[\g\,(\j^{(1)}+\j^{(2)}+\j^{(3)})^3\big]$. Normalizing this to a sum of standard QM pairwise probabilities, we define for the simplified system~\eqref{e:3slitss}:
\begin{equation}
  \kappa(1,2,3)
   := \frac{ \g\,(\j^{(1)}+\j^{(2)}+\j^{(3)})^3 }
           { \big( P_2(1,2)+P_2(1,3)+P_2(2,3) \big)^{3/2} },
 \label{e:kappa3}
\end{equation}
where the fractional power in the denominator insures that the quotient is $\j$-scale invariant\footnote{If the state-space is homogeneous, the $\j$'s are {\em\/sections\/} over it with definite degrees of homogeneity. The definition~\eqref{e:kappa3} then insures that $\kappa(1,2,3)$ has well-defined (scale-invariant) values over the whole state-space.}.
 The strictly trilinear interference quotient
\begin{equation}
  \widehat\kappa(1,2,3)
   := \frac{ \g\,\j^{(1)}\,\j^{(2)}\,\j^{(3)} }
           { \big( P_2(1,2)+P_2(1,3)+P_2(2,3) \big)^{3/2} },
 \label{e:kappa3t}
\end{equation}
shares the features of $\kappa(1,2,3)$ but has a smaller amplitude.
 They are both periodic in time, with the same frequency, $\w=\sqrt{\w_0^{~2}{+}2\g c_2}$ --- which is a consequence of the simplification in the toy model~\eqref{e:3slitss}, where each of the three state-spaces was collapsed to a single (ground) state.\footnote{Including additional states introduces multiple frequencies, and thereby amplitude modulation and ``beats,'' such as reported in the nonlinear optics study~\cite{Namdar:2021czo}.} Nevertheless, owing to the non-simple harmonic nature of the Jacobi elliptic functions~\eqref{e:psi123}, even this drastically simplified toy model exhibits a non-simple harmonic structure visible in Figure~\ref{f:kappa}.
It is worth noting that the overall amplitudes of $\kappa(1,2,3)$ and $\widehat\kappa(1,2,3)$  are both proportional to $\g$ but scale about one order of magnitude smaller; $\widehat\kappa(1,2,3)$ is also proportional to $k^2\w^2$.

Reparametrizing a branch of the solutions~\eqref{e:psi123} as
\begin{subequations}
 \label{e:Psi123}
\begin{alignat}9
 \j^{(1)}(t)
 &= A\;\mathrm{sn}\big(\w\,t\,,-(\tfrac{A\g}{\w})^2\big),\qquad
 \j^{(2)}(t)
  = A\;\mathrm{cn}\big(\w\,t\,,-(\tfrac{A\g}{\w})^2\big),\\
 \j^{(3)}(t)
 &= \tfrac\w\g\,\mathrm{dn}\big(\w\,t\,,-(\tfrac{A\g}{\w})^2\big)-\tfrac{\w_0}\g, \label{e:Psi3}
\end{alignat}
\end{subequations}
the initial $t=0$ state triple is $\big(0,A,\tfrac{\w-\w_0}\g\big)$, the {\em\/standard\/} normalization of which would require $A^2\,{=}\,1{-}\big(\tfrac{\w-\w_0}\g\big)^2$. Reality of $\j^{(i)}(t)$ than requires that $|\w{-}\w_0|<\g$. The {\em\/standard\/} evolution probability of {\em\/not decaying\/} from the initial state to some later time is then
\begin{equation}
\begin{aligned}
 P(t|&0)
 =\Big(\sum_i \j^{(i)}(t)^{\textstyle*}\,\j^{(i)}(0)\Big)^2,\\
 &=\Big(\big[1{-}\big(\tfrac{\w{-}\w_0}\g\big)^2\big]\,
    \mathrm{cn}\big(\w t;\tfrac{(\w{-}\w_0)^2{-}\g^2}{\w^2}\big)
   +\tfrac{\w -\w_0}\g\big[\tfrac\w\g\,\mathrm{dn}\big(\w t;\tfrac{(\w{-}\w_0)^2{-}\g^2}{\w^2}\big)-\tfrac{\w_0}\g\big]\Big)^2,
\end{aligned}
\end{equation}
which indeed satisfies $\lim_{t\to0^+}P(t|0)=1$ and $\dot{P}(t|0)<0$. Expanding to $O(\g)$ yields
\begin{equation}
  P(t|0)\approx\big(\xi^2{+}(1{-}\xi^2)\cos(\w t)\big)^2
   -\xi(1{-}\xi)^2\frac{\g}{\w_0}
       \big(\xi^2{+}(1{-}\xi^2)\cos(\w t)\big) \sin^2(\w t),
\end{equation}
where $\xi=c_2/\w_0$ is the 2nd integration constant in units of $\w_0$, restricted to $|\xi{-}\tfrac\g{2\w_0}|<1$ by the reality of $\j^{(i)}(t)$, as above. The results of this very crude toy-model qualitatively remind of the more specific results in Nambu Quantum Mechanics~\cite{Minic:2020zjb,nabin}.

\section{Phenomenology: non-linear Talbot effect}
\label{s:Talbot}
Another robust probe of coherent interference effects involving matter waves utilizes the Talbot effect \cite{talbot}, in which, in the classic
set-up of a plane wave diffracting on the diffraction grating, the image of
the grating appears at regular distances (Talbot length), and also, self images (Talbot images) appear at fractions of
the Talbot length \cite{talbot1}. The Talbot effect can be observed in the quantum context as well (see \cite{talbot2}, \cite{talbotq}.)
Finally, there is a non-linear version of the Talbot effect that has been explored numerically by using the non-linear
Schrodinger equation \cite{nltalbot}.

What is of interest to us is that the non-linear Talbot effect does have very noticable signatures \cite{nltalbot}.
The model essentially involves the following state dependent change of the Hamiltonian in the canonical Schrodinger equation
\be
H \to H + |\psi|^2
\ee
What is interesting here is that this change has been discussed in our previous paper on ``gravitization of the quantum''
\cite{Berglund:2022skk}.
Also, the same effective non-linear equation can  be obtained from our toy model (discussed in the previous section)
by replacing (to leading order in $\gamma$) $\psi_3$ with $\psi_3 \to  |\psi|^2$ (where the real and imaginary components of $\psi$
are precisely $\psi_1$ and $\psi_2$)
\be\label{eq:NLSE}
i \frac{\partial{\psi}}{\partial t} = H \psi + \gamma |\psi|^2 \psi.
\ee
This is the non-linear Schrodinger equation considered in \cite{nltalbot} (after reabsorbing $\gamma$ via $\psi \to \sqrt{\gamma} \psi$).~\footnote{We emphasize that non-linear modifications to the Schrodinger equation while maintaining a fixed Hilbert space structure for quantum mechanics are qualitatively different than what we are proposing.}
Note that in \cite{nltalbot} the Hamiltonian is chosen to be that of a free particle in one dimension.

\begin{figure}
\centering
\begin{subfigure}{.5\textwidth}
  \centering
  \includegraphics[scale=0.5]{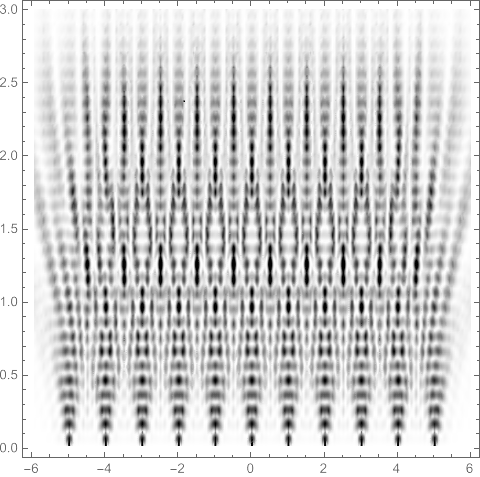}
  \caption{Without non-linear modification}
  \label{fig:sub1}
\end{subfigure}%
\begin{subfigure}{.5\textwidth}
  \centering
  \includegraphics[scale=0.5]{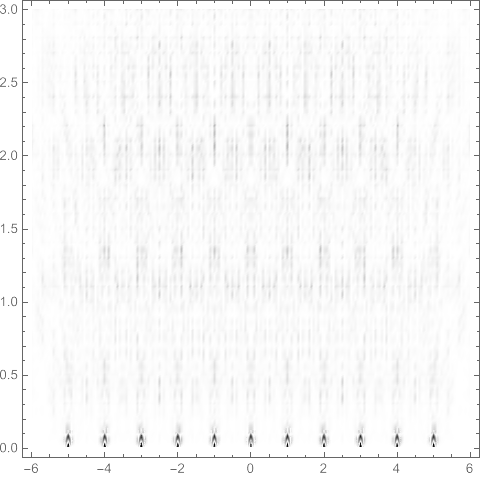}
  \caption{With non-linear modification}
  \label{fig:sub2}
\end{subfigure}
\caption{Numerically generated partial, near-slit Talbot carpets for the unmodified Schrodinger equation as well as the non-linear Schrodinger equation in Eq.~\ref{eq:NLSE}. The slits generating the pattern are at $y=0$, the size of the non-linear coefficient was chosen for illustration, and the probability density has been scaled by $y^{1/2}$ to visually enhance the pattern for large $y$.}
\label{fig:NLT}
\end{figure}

The dramatic signature of the non-linear Talbot effect is illustrated in Figure~4 of \cite{nltalbot} as well as in Figure~\ref{fig:NLT}.
In particular, the claim is the fundamental non-linearity is
manifested in the ``destruction'' of the self-similar structure
of the Talbot carpet. We would claim that this 
destruction of the fractal structure of 
the Talbot carpet of (quantum) matter waves 
happens in the context of quantum gravity,
viewed as
``gravitization of the quantum'', as discussed in this paper.~\footnote{In the context of the non-linear Talbot effect in
principle we have not only intrinsic triple but also
higher order interference effects. These can
be modeled by including 
$\psi_4$, $\psi_5$ etc and the
corresponding extra Nambu-like equations that 
naturally extend out current toy model.}

{Let us discuss the possible parameter space. In our toy model $H$ is of the order of eV, and $\gamma$, by the argument in section~\ref{s:scale}, should be of the order of 1 TeV. Is this observable? The claim 
would be ``yes'', because of the non-linearity of the effect. 
Note that the effective scale of $10^{-19}m$ that corresponds to
the Higgs scale is probed in gravitational interferometry.
So gravitational interferometry  \cite{Holometer:2016ipr}
could be one important probe.
(Note that, unlike \cite{Holometer:2016ipr},
we propose a probe of quantum gravity based
on tests of the foundations of quantum theory.)

However, we should also ask:
Does the cosmological constant scale of $10^{-3}$ eV enter this discussion? This is the
scale of $10^{-4}$m which is usually the characteristic scale associated with the ``thickness'' of the extra dimension in that particular approach to the hierarchy problem. This scale might
be probed in the AMO interferometry experiments. One argument is as follows: In the context of non-linear deformations of
the canonical quantum theory there exist very tight bounds on
the relevant deformation parameters. For example, neutron interferometry 
\cite{zeilinger} tightly constrains logarithmic 
($b \psi \log{a |\psi|^2}$, with 
$b< 10^{-15}$ eV)
additions to the Sch\"{o}dinger equation of canonical quantum theory \cite{bb}, 
and cosmology might put even
stronger constraints equivalent to the energy scale associated with the Hubble scale $b_{CMB} < 10^{-34}$ eV (and a couple of orders of
magnitude smaller $b_{now} < 10^{-37}$ eV) \cite{Georg:2018dza} .
Now, this pretty much rules out such deformations of the canonical quantum theory based on a single Hilbert space. But those are highly constrained by the rigid geometry of quantum theory, as we have emphasized earlier. However, in the context of multiple Hilbert spaces
that are observer-dependent, such bounds can be weakened
considerably by factors associated with large numbers of the order $10^{31}$ discussed in section~\ref{s:scale}, and the effective bound could be equivalent to the
energy scale associated with the cosmological constant $10^{-3} eV$ (or a couple of orders of magnitude 
smaller, $10^{-6}$ eV). This would consistent with the
current constraints on triple interference, reviewed in section~\ref{s:3+ference}, as well as constraints from Nambu quantum theory \cite{nabin}. (By comparing our toy model to the constraints from \cite{nabin} our $\gamma$ parameter is less than $10^{-2}$ eV, provided that the Hamiltonian $\omega_0$ is of the order of eV, as discussed in the next section.)
Future experiments could improve this by a few orders of magnitude.}

\section{Phenomenology: Experimental tests with AMO}
\label{s:Exp}

The degree of quantum control in state preparations, operations, and readout of individual atoms has been recognized with the 2012 Nobel prize in Physics \cite{Nobel2012}  
and has led to many promising applications including quantum computing, simulation, and precision sensing. Such a level of quantum control can be used to test the degree to which the Born rule in massive quantum systems is shown to hold in the low-energy limit. For instance, ion systems have degrees of freedom that can be effectively isolated and form qubits useful for quantum information processing or quantum simulation. High fidelity gate operations performed on arrays of single trapped ions, where single qubit gate fidelities have approached 99.9999 percent \cite{singlequbit1} and 2 qubit gates have recently achieved 99.9 percent \cite{2qubit} fidelity.
Considering Eq.~(35) of Ref. \cite{Minic:2020zjb}, for a two-level system interacting with a harmonic perturbation e.g. for a transition between two hyperfine states, the probability of inducing a Rabi spin flip is modulated by a pre-factor $\cos^2{\xi}+k^2\sin^2{\xi}$, where $0\leq k^2<1$.
\begin{equation}
 P(|0\rangle \rightarrow |1\rangle)=\left[\cos^2{\xi}+k^2\sin^2{\xi}\right]\Big(\frac{\Omega^2}{\Omega^2+\Delta^2}\Big)\sin^2({\sqrt{\Omega^2+\Delta^2}\tau_p/2})
 \end{equation}
 where $\Omega$ is the Rabi frequency, $\Delta$ is the detuning of the laser from the atomic resonance, and $\tau_p$ is the pulse duration.  For $\Delta=0$, when $\Omega \tau_p = \pi$, this is known as a ``$\pi$'' pulse, perfectly inverting the population. For nonzero detuning the effective Rabi frequency $\Omega_{\mathrm{eff}}=\sqrt{\Omega^2+\Delta^2}$ increases but the probability drops below 1. 
The demonstration of near-perfect $\pi$ pulses allows us to place bounds on the prefactor $\left[\cos^2{\xi}+k^2\sin^2{\xi}\right]$. 
(Note that $k \sin{\xi}$ maps to the $\gamma$ parameter in our toy model, and given the bounds found in \cite{nabin}, the
ratio of $\gamma$ to the characteristic energy $\omega_0$ in our toy model is currently weakly bound by $10^{-2}$.)
Such a bound may be used to place limits on violation of the Born rule in general.  

Though these manipulations on quantum bits can test the validity of quantum theory, a triple interference experiment, for example based on a Mach-Zehnder-type light-pulse atom interferometer, could be a useful tool to search for the presence of the particular triple interference measures of the type studied in our toy model $\kappa$ and $\hat{\kappa}$ in Eqs.~(\ref{e:kappa3}) and~(\ref{e:kappa3t}). A detailed analysis of the expected interference fringes for such a setup is left as a subject for future work.  A potentially more powerful probe of non-linear modifications of the Born rule involves the intrinsically multi-slit interference phenomenon of the near-field Talbot effect \cite{talbot}.
Atomic interferometry has also been used to examine the Talbot effect \cite{Sleator05}. The nonlinear Talbot effect manifests as a significant reduction in the periodic pattern referred to as the Talbot carpet. In Ref. \cite{Sleator05}, the authors studied the period of the Talbot carpet at a variety of fractional multiples of the Talbot length following a grating.  In principle, from the existence of this pattern, one could set a bound on the parameter $\gamma$ in our toy model. 
Looking forward, future experiments involving matter-wave diffraction of nanoparticles of masses up to $10^7$ atomic mass units can also be used to probe the Talbot effect and search for nonlinearities.

\section{Concluding thoughts}
\label{s:Coda}
In this paper we have discussed a new approach to the problem of quantum gravity in
which the quantum mechanical structures that are traditionally fixed, such as the Fubini-
Study metric in the Hilbert space of states, become dynamical and so implement the idea
of gravitizing the quantum. 
After reviewing the concept of triple, and higher
order, interference in
section~\ref{s:3+ference}, we discussed triple interference in the context of
non-linear optics as a well-known analogy in section~\ref{s:NLO}, and in section~\ref{s:3nterference} we proposed a concrete model of triple
interference in the context of a gravitized quantum theory.
Given various arguments for the robustness of the Born rule, reviewed
in section~\ref{s:dragons}, we provided generic answers to
various criticisms relating to possible modifications of quantum theory. 
In section~\ref{s:GQ=QG},
we tied this proposal to both a theoretical proposal for quantum gravity as
well as a geometric view of quantum theory and in section~\ref{s:scale} we discussed the scale of triple interference involving
quantum gravity.
In particular, in section~\ref{s:toy} we discussed a specific test of this new
approach to quantum gravity using triple interference in a varying gravitational field,
driven by a profound analogy with recent triple-path interference experiments
performed in the context of non-linear optics. We emphasized that the triple interference
experiment in a varying gravitational field would deeply influence the present understanding
of the kinematics of quantum gravity. Another proposal referred to a conformally deformed
FS (Fubini-Study) metric, and the red-shift experiments based on the optical lattice atomic
clocks was mentioned in section~\ref{s:GQ=QG}. Finally, in sections~\ref{s:Talbot} and~\ref{s:Exp} we discussed the non-linear Talbot effect as an even more striking probe of
quantum gravity viewed as gravitization of the geometry of quantum theory. 
We plan to
return to a more detailed treatment of the actual experimental arrangement for this and
other effects based on intrinsic multiple interference in our future work.

\paragraph{Acknowledgments:} We thank Sougato Bose, Laurent Freidel, Jerzy Kowalski-Glikman, Rob Leigh, Anupam Mazumdar, 
Holger M\"{u}ller and Tatsu Takeuchi for
inspiring discussions. 
PB would like to thank the CERN Theory Group
for their hospitality over the past several years. TH is grateful to the Department of Physics,
University of Maryland, College Park MD, and the Physics Department of the Faculty of
Natural Sciences of the University of Novi Sad, Serbia, for the recurring hospitality and resources.
D. Minic is grateful to Perimeter Institute for its hospitality and support. The work of PB and D. Mattingly
is supported in part by the Department of Energy grant DE-SC0020220.
D. Minic thanks the Julian Schwinger Foundation and the U.S. Department
of Energy under contract DE-SC0020262 for support. 
{A. Geraci is partially supported by
NSF grant nos. PHY-1505994, 1806686, the Heising-Simons
Foundation, the J. Templeton Foundation, and the office of
Naval Research grant no.417315//N00014-18-1-2370.}
This work contributes to the European Union COST Action CA18108 {\it Quantum gravity phenomenology in the multi-messenger approach}.



\begin{thebibliography}{100}
\frenchspacing
\raggedright

%
\bibitem{Berglund:2022skk}
P.~Berglund, T.~H\"ubsch, D.~Mattingly and D.~Minic,
``Gravitizing the Quantum,''
Int. J. Mod. Phys. {\bf31D} (2022), 2242024,
[arXiv:2203.17137 [gr-qc]].

\bibitem{Berglund:2022qcc}
P.~Berglund, L.~Freidel, T.~H{\"u}bsch, J.~Kowalski-Glikman, R.~G.~Leigh, D.~Mattingly and D.~Minic,
``Infrared Properties of Quantum Gravity: UV/IR Mixing, Gravitizing the Quantum --- Theory and Observation,''
[arXiv:2202.06890 [hep-th]].


%
\bibitem{Freidel:2015uug} %
L.~Freidel, R.~G. Leigh, and D.~Minic, 
Int. J. Mod. Phys.\,D
  {\bf 24} (12), (2015) 1544028.
%
\bibitem{Freidel:2016pls} %
L.~Freidel, R.~G. Leigh, and D.~Minic, 
Phys. Rev.\,D
  {\bf 94} (10), (2016) 104052.
%
\bibitem{Freidel:2013zga} %
L.~Freidel, R.~G. Leigh, and D.~Minic, 
 Phys. Lett.\,B
 {\bf 730} (2014) 302--306. 
%
\bibitem{Freidel:2014qna} %
L.~Freidel, R.~G. Leigh, and D.~Minic, 
Int. J. Mod. Phys.\,D
{\bf 23} (12), (2014) 1442006. 
%
\bibitem{Freidel:2015pka} %
L.~Freidel, R.~G. Leigh, and D.~Minic, 
JHEP
{\bf 06} (2015) 006.
%
\bibitem{Freidel:2017xsi} %
L.~Freidel, R.~G. Leigh, and D.~Minic,
 J. Phys. Conf. Ser. {\bf  804} (1) (2017) 012032.
%
\bibitem{Freidel:2018apz} %
L.~Freidel, J.~Kowalski-Glikman, R.~G. Leigh, and D.~Minic,
 Phys.\ Rev.\,D {\bf99} (6) (2019) 066011.
%
\bibitem{Freidel:2019jor}
L.~Freidel, R.~G.~Leigh and D.~Minic,
Int.\ J.\ Mod.\ Phys.\,A {\bf 34} (28), (2019) 1941004.
%
\bibitem{Minic:2020oho}
D.~Minic,
[arXiv:2003.00318 [hep-th]].
%
\bibitem{Freidel:2017wst} %
L.~Freidel, R.~G. Leigh, and D.~Minic, 
 JHEP {\bf 09} (2017) 060. 
%
\bibitem{Freidel:2017nhg} %
L.~Freidel, R.~G. Leigh, and D.~Minic, 
Phys. Rev.\,D {\bf 96} (6) (2017) 066003.

%
\bibitem{Freidel:2021wpl}
L.~Freidel, J.~Kowalski-Glikman, R.~G.~Leigh and D.~Minic,
Int. J. Mod. Phys. D \textbf{30}, no.14, 2141002 (2021)
[arXiv:2104.00802 [gr-qc]].

%
\bibitem{Gunaydin:2013nqa}
M.~G\"unaydin and D.~Minic,
Fortsch. Phys. \textbf{61}, 873-892 (2013)
[arXiv:1304.0410 [hep-th]].


\bibitem{Berglund:2020qcu}
P.~Berglund, T.~H\"ubsch and D.~Minic,
LHEP \textbf{2021}, 186 (2021)
[arXiv:2010.15610 [hep-th]].


\bibitem{Minic:2002pd}
D.~Minic and H.~C.~Tze,
Phys. Lett. B \textbf{536}, 305-314 (2002)
[arXiv:hep-th/0202173 [hep-th]].



\bibitem{Minic:2003en}
D.~Minic and H.~C.~Tze,
Phys. Rev. D \textbf{68}, 061501 (2003)
[arXiv:hep-th/0305193 [hep-th]].


%
\bibitem{Minic:2004rj}
D.~Minic and H.~C.~Tze,
Contribution to QTS-3 conference proceedings, 159-166,
[arXiv:hep-th/0401028 [hep-th]].




%
\bibitem{Jejjala:2007rn}
V.~Jejjala, M.~Kavic and D.~Minic,
Int. J. Mod. Phys. A \textbf{22}, 3317-3405 (2007)
[arXiv:0706.2252 [hep-th]].




\bibitem{Minic:2020zjb}
D.~Minic, T.~Takeuchi and C.~H.~Tze,
Phys. Rev. D \textbf{104}, no.5, L051301 (2021)
[arXiv:2012.06583 [hep-ph]].

\bibitem{Raju} 
K. Papadodimas, and S. Raju, 
Phys. Rev. D, vol. 89, no. 8, p. 86010, 2014, doi: 10.1103/Phys- RevD.89.086010.

\bibitem{penrose}
R.~Penrose, 
Foundations of Physics, \textbf{44}, 557 (2014).



\bibitem{Hardy:2001jk}
L.~Hardy,
``Quantum theory from five reasonable axioms,''
[arXiv:quant-ph/0101012 [quant-ph]]; 
``Probability theories with dynamic causal structure: A New framework for quantum gravity,''
[arXiv:gr-qc/0509120 [gr-qc]].

\bibitem{Norton}
J.~Norton, 
Reports on Progress in Physics, Vol. \textbf{56}, No. 7 (1993).

%


%
\bibitem{Addazi:2021xuf}
A.~Addazi, J.~Alvarez-Muniz, R.~A.~Batista, G.~Amelino-Camelia, V.~Antonelli, M.~Arzano, M.~Asorey, J.~L.~Atteia, S.~Bahamonde and F.~Bajardi, \textit{et al.}
[arXiv:2111.05659 [hep-ph]].

%
\bibitem{Bose:2017nin}
S.~Bose, A.~Mazumdar, G.~W.~Morley, H.~Ulbricht, M.~Toro\v{s}, M.~Paternostro, A.~Geraci, P.~Barker, M.~S.~Kim and G.~Milburn,
Phys. Rev. Lett. \textbf{119}, no.24, 240401 (2017)
[arXiv:1707.06050 [quant-ph]].

%
\bibitem{Marletto:2017kzi}
C.~Marletto and V.~Vedral,
Phys. Rev. Lett. \textbf{119}, no.24, 240402 (2017)
[arXiv:1707.06036 [quant-ph]].


\bibitem{Belenchia:2018szb}
A.~Belenchia, R.~M.~Wald, F.~Giacomini, E.~Castro-Ruiz, \v{C}.~Brukner and M.~Aspelmeyer,
Phys. Rev. D \textbf{98}, no.12, 126009 (2018)
[arXiv:1807.07015 [quant-ph]].

\bibitem{Carney:2022dku}
D.~Carney, Y.~Chen, A.~Geraci, H.~M\"uller, C.~D.~Panda, P.~C.~E.~Stamp and J.~M.~Taylor,
``Snowmass 2021 White Paper: Tabletop experiments for infrared quantum gravity,''
[arXiv:2203.11846 [gr-qc]].



%





%

\bibitem{Sorkin:1994dt}
R.~D.~Sorkin,
Mod. Phys. Lett. A \textbf{9}, 3119-3128 (1994)
[arXiv:gr-qc/9401003 [gr-qc]].

\bibitem{Huber:2021xpx}
P.~Huber, H.~Minakata, D.~Minic, R.~Pestes and T.~Takeuchi,
[arXiv:2105.14061 [hep-ph]].

\bibitem{Helou and Chen}
B.~Helou and Y.~Chen,
Journal of Physics: Conference Series, \textbf{880}, 1, (2017), arXiv:1709.06639 [quant-ph].

\bibitem{Sinha}
U.~Sinha, C.~Couteau, T.~Jennewein, R.~Laflamme,
and G. Weihs, Science \textbf{329}, 418 (2010), arXiv:1007.4193[quant-ph].

\bibitem{Park}
D.~K.~Park, O.~Moussa, and R.~Laflamme, New J. Phys. \textbf{14}, 113025 (2012), arXiv:1207.2321 [quant-ph].
Phys. Rev. Lett. 126, 190401.

\bibitem{Lutz1}
M.-O.~Pleinert, A.~Rueda, E.~Lutz, and J.~von Zanthier, 
Phys. Rev. Lett. \textbf{126}, 190401.

\bibitem{Lutz2}
M.-O.~Pleinert, J.~von Zanthier and E.~Lutz, 
Phys. Rev. Research \textbf{2}, 012051(R) (2020).

\bibitem{nonclassical} R. Sawant, J. Samuel, A. Sinha, S. Sinha, and U. Sinha, 
Phys. Rev. Lett., vol. 113, no. 12, p. 120406, 2014, doi: 10.1103/ PhysRevLett.113.120406.

%
\bibitem{Namdar:2021czo}
P.~Namdar, P.~K.~Jenke, I.~A.~Calafell, A.~Trenti, M.~Radonji\'c, B.~Daki\'c, P.~Walther and L.~A.~Rozema,
``Experimental Higher-Order Interference in a Nonlinear Triple Slit,''
[arXiv:2112.06965 [quant-ph]].

\bibitem{WW}
Whittaker, E. and Watson, G. (1996),
A Course of Modern Analysis (4th ed., Cambridge Mathematical Library). 
Cambridge: Cambridge University Press. doi:10.1017/CBO9780511608759

\bibitem{mcp}
K. S. Thorne and R. D. Blandford, {\it Modern Classical Physics}, Princeton, 2017.


\bibitem{bornrule}
M. Born, Zeitschrift fur Physik 37, 863-867 (1926).

\bibitem{earman}
J. Earman, {\it The Status of the Born Rule and the Role of  Gleason's Theorem and Its Generalizations:  How the Leopard Got Its Spots and Other Just-So Stories}, http://philsci-archive.pitt.edu/20792/, and references therein.

\bibitem{island}
S. Aaronson, {\it Is Quantum Mechanics An Island In Theory Space? },  arXiv:quant-ph/0401062.

\bibitem{gleason}
A. M. Gleason, Journal of Mathematics and Mechanics 6: 855-893, 1957.




\bibitem{mgm}
L. Masanes, T. D. Galley, M. P. Müller, Nat. Commun. 10, 1361 (2019), and references therein.


\bibitem{weinbergQFT}
S. Weinberg, {\it The Quantum Theory of Fields, volume 1: Foundations}, Cambridge University Press, 1995. 

\bibitem{Donnelly:2016auv}
W.~Donnelly and L.~Freidel,
JHEP \textbf{09}, 102 (2016)
[arXiv:1601.04744 [hep-th]].

\bibitem{Marolf:2020xie}
D.~Marolf and H.~Maxfield,
JHEP \textbf{08}, 044 (2020)
[arXiv:2002.08950 [hep-th]].


%
\bibitem{Witten:1985cc}
E.~Witten,
Nucl. Phys. B \textbf{268}, 253-294 (1986)

%
\bibitem{Strominger:1987bn}
A.~Strominger,
Phys. Lett. B \textbf{187}, 295 (1987)


%
\bibitem{Zwiebach:1992ie}
B.~Zwiebach,
Nucl. Phys. B \textbf{390}, 33-152 (1993)
[arXiv:hep-th/9206084 [hep-th]].

\bibitem{SFT1} C. de Lacroix, H. Erbin, S. P. Kashyap, A. Sen, and M. Verma, ``Closed Superstring Field Theory and its Applications,'' Int. J. Mod. Phys., vol. 32, no. 28n29, p. 1730021, 2017, doi: 10.1142/S0217751X17300216.

\bibitem{SFT2} T. Erler, 
Phys. Rept., vol. 980, pp. 1–95, 2022, doi: 10.1016/j.physrep.2022.06.004.

\bibitem{SFT3} T. Erler, 
Phys. Rept., vol. 851, pp. 1–36, 2020, doi: 10.1016/j.physrep.2020.01.003.

\bibitem{SFT4} H. Erbin, String Field Theory: A Modern Introduction, vol. 980, 2021.

\bibitem{Polchinski:1998rq}
  J.~Polchinski,
  {\it String theory}, 
Cambridge University Press, 1998.
M.~B.~Green, J.~H.~Schwarz and E.~Witten, {\it Superstring theory}, Cambridge University Press, 1987.








%

%
\bibitem{Barnum:2014ysa}
H.~Barnum, M.~P.~M\"uller and C.~Ududec,
New J. Phys. \textbf{16}, no.12, 123029 (2014)
[arXiv:1403.4147 [quant-ph]].


\bibitem{tripleint}
C.~Ududec, H.~Barnum, J.~Emerson, 
Foundations of Physics --- Advances in Quantum Foundations issue,  arXiv:0909.4787v1 [quant-ph].


\bibitem{twohilbert}
S.~Bunk, L.~Muller  and R.~J.~Szabo,
  Lett.\ Math.\ Phys.\  {\bf 109}, no. 8, 1827 (2019);
  V.~Balasubramanian, J.~de Boer and D.~Minic,
  gr-qc/0211003;
L.~Smolin,
  J.\ Math.\ Phys.\  {\bf 36}, 6417 (1995);
L.~Crane,
  J.\ Math.\ Phys.\  {\bf 36}, 6180 (1995).
D.~Minic,
  Phys.\ Lett.\ B {\bf 442}, 102 (1998);
T.~Banks and W.~Fischler,
  ``M theory observables for cosmological space-times,''
  hep-th/0102077.


%
\bibitem{Ashtekar:1997ud}
A.~Ashtekar and T.~A.~Schilling,
[arXiv:gr-qc/9706069 [gr-qc]].








\bibitem{aa}
J.~Anandan and Y.~Aharonov, Phys. Rev. Lett. \textbf{65}, 1697 (1990);
J.~Anandan, Found. Phys. 21, 1265 (1991).

\bibitem{katori}
M.~Takamoto, I.~Ushijima, N.~Ohmae et al., 
Nat. Photonics \textbf{14}, 411–415 (2020).



\bibitem{Freidel:2022ryr}
L.~Freidel, J.~Kowalski-Glikman, R.~G.~Leigh and D.~Minic,
``The Vacuum Energy Density and Gravitational Entropy,''
[arXiv:2212.00901 [hep-th]].

%
\bibitem{Berglund:2022qsb}
P.~Berglund, T.~H\"ubsch and D.~Minic,
``On de Sitter Spacetime and String Theory,''
[arXiv:2212.06086 [hep-th]], accepted for publication in IJMPD.


\bibitem{extradim}
I. Antoniadis, Phys. Lett. B 246, 377 (1990); 
N. Arkani-Hamed, S. Dimopoulos and G. R. Dvali, Phys. Lett. B 429, 263 (1998);
L. Randall and R. Sundrum, Phys. Rev. Lett. 83, 3370 (1999); Phys. Rev. Lett. 83, 4690 (1999).




\bibitem{talbot}
H. F. Talbot, The London, Edinburgh, and Dublin Philosophical Magazine and Journal of Science. 9 (56): 401–407 (1836).

\bibitem{talbot1}
Lord Rayleigh, The London, Edinburgh, and Dublin Philosophical Magazine and Journal of Science. 11 (67): 196–205 (1881). 

\bibitem{talbot2}
M. Berry et al, Physics World, June 2001, and references therein.

\bibitem{talbotq}
M. Chapman, et al,
Physical Review A. 51 (1): R14–R17 (1995).

\bibitem{nltalbot}
Y. Zhang, et al, Physical Review E. 89 (3): 032902 (2014); arXiv:1402.3017, and references therein.


%
\bibitem{Holometer:2016ipr}
A.~Chou \textit{et al.} [Holometer],
Class. Quant. Grav. \textbf{34}, no.6, 065005 (2017)
[arXiv:1611.08265 [physics.ins-det]].



\bibitem{zeilinger}
R. G\"{a}hler, A.G.Klein, A. Zeilinger, Phys. Rev. A23, 1611 (1981).

\bibitem{bb}
I. Bialynicki-Birula and J. Mycielski, Ann. of Phys. 100, 62 (1976).
%
\bibitem{Georg:2018dza}
J.~Georg and C.~Rosenzweig,
JCAP \textbf{01}, 028 (2020)
[arXiv:1804.07305 [gr-qc]].


\bibitem{nabin}
N. Bhatta, D. Minic and T. Takeuchi,
{\it Constraints on Nambu Quantum Mechanics from Oscillation Phenomena}, in preparation.


\bibitem{Nobel2012}
Awarded to Serge Haroche and David J. Wineland ``for ground-breaking experimental methods that enable measuring and manipulation of individual quantum systems'',
https://www.nobelprize.org/prizes/physics/2012/summary/ .

\bibitem{singlequbit1}
T. P. Harty, D. T. C. Allcock, C. J. Ballance, L. Guidoni, H. A. Janacek, N. M. Linke, D. N. Stacey and D. M. Lucas,
Phys. Rev. Lett. 113, 220501 (2014).

\bibitem{2qubit}
C. J. Ballance, T. P. Harty, N. M. Linke, M. A. Sepiol, and D. M. Lucas, Phys. Rev. Lett. 117, 060504 (2016).
  
\bibitem{Sleator05}
A. Turlapov, A. Tonyushkin, and T. Sleator,
Phys. Rev. A 71, 043612 (2005).

\end{thebibliography}
\end{document}